\documentclass[aps,twocolumn,pre]{revtex4-1}

\usepackage{graphicx}
\usepackage{amsmath,mathtools}
\usepackage{color}

\renewcommand{\i}{\mathrm{i}}

\usepackage[hidelinks]{hyperref}

\begin{document}

\title{Efficiency fluctuations of a heat engine with noise induced  quantum coherences}

\author{Manash Jyoti Sarmah}

\author{Himangshu Prabal Goswami}
\email{hpg@gauhati.ac.in}
\affiliation{Department of Chemistry, Gauhati University, Jalukbari, Guwahati-781014, Assam, India}
\date{\today}

\begin{abstract} 
We analyze the efficiency fluctuations of a coherent quantum heat engine coupled to a unimodal cavity using a standard full-counting statistics procedure.
The engine's most likely efficiency obtained by computing the large-deviation function corresponds to the quantum efficiency obtained by defining a useful work obtainable from a steady-state fluctuation theorem. The most likely efficiency is  independent of the noise induced coherences. The ratio between coherent and incoherent efficiency cumulants is found to be constant and independent of the stochastic efficiency, which we prove analytically. We also numerically demonstrate that treating the efficiency as a stochastic variable allows the enhancement of constancy only in the presence of coherences. At the engine's most likely efficiency, there is no suppression of the second efficiency cumulant leading to a robust constancy. We also report the existence of a lower bound on the second efficiency cumulant.
\end{abstract}

\maketitle

\section{\label{sec-intro}
Introduction
}

At the quantum level, thermodynamic observables exhibit randomness due to the presence of significant thermal fluctuations, alongside additional quantum fluctuations when temperatures are sufficiently low. 
These fluctuations lead to variations not only in the energy, work, and heat but also in the efficiency of a system during any energy, particle, or information transfer processes. In classical thermodynamics, the efficiency of a heat engine is bounded by the Carnot efficiency, which depends solely on the temperature of the reservoirs. However, in quantum systems,  fluctuations introduce variability around this ideal efficiency. Since both heat and work are fluctuating in the quantum realm, a stochastic efficiency emerges. Efficiency fluctuations are related to the dissipative nature of non-equilibrium quantum systems. Analyzing these fluctuations, insightful access into the quantum thermodynamic aspect of non-equilibrium phenomena is gained. This allows for the development of strategies to mitigate the impact of fluctuations on controllable quantum energy transfer devices. Recent studies have extensively investigated the theoretical realm of efficiency fluctuations in classical heat engines\cite{Gingrich_2014Efficiency_large_deviations, PhysRevE.90.052145_Universal_theory_of_efficiency, PhysRevE.93.052123_eff_flu_small_machines, PhysRevLett.114.050601Efficiency_Statistics_at_All, PhysRevLett.122.140601_eff_flu_micro_mac, Verley2014_unlikely_Carnot_efficiency}. These studies revealed that the stochastic efficiency of a Carnot engine can surpass the Carnot bound\cite{Verley2014_unlikely_Carnot_efficiency}. Interestingly, this upper limit has also been identified as the least probable outcome over extended periods\cite{Verley2014_unlikely_Carnot_efficiency}. \par Experimental evidence supporting theoretical predictions has been observed in a stochastic engine utilizing an optically trapped colloidal particle\cite{martinez2016_brownian_carnot_engine}. Exploring the quantum domain, studies have extensively investigated the work distributions of driven oscillators and two-level systems\cite{PhysRevE.77.021128_non_eq_work_dis, PhysRevLett.101.070403_Trapped_Cold_atom_emp, PhysRevE.78.011115_forced_qun_osc, PhysRevB.87.060508_fluc_dri_q_sys, PhysRevLett.111.093602_w_didd_2_l_sys}, utilizing both theoretical models and also experimental setups such as ion-trap\cite{An2015_ion_trap}, NMR\cite{PhysRevLett.113.140601ExperimentalReconstructionWorkDistribution}, and cold-atom systems\cite{Cerisola2017_colsAtom_setups}. Moreover, studies have analyzed the efficiency fluctuations of a quantum thermoelectric junction using full-counting statistics\cite{PhysRevB.91.115417_eff_fluc_q_thermo_ele_devices}. The efficiency of a SWAP quantum engine has also been numerically assessed through simulations of calorimetric measurements\cite{Campisi_2015_egs_swap}. Small machines display behavior similar to their macroscopic system counterparts in the sense that these convert a mean input flux into a mean output flux while adhering to the reversible efficiency as prescribed by the second law of thermodynamics\cite{book_thermodynmics}. However, these input and output fluxes fluctuate with root mean squares that may exceed their average values. These fluctuations are governed by universal fluctuation relations, which ultimately uphold the second law when ensemble-averaged\cite{Sinitsyn_2011_FT_1, Campisi_2014_FT2, e20090635_FT3}. Consequently, the efficiency ($\eta$) of a machine during a single duration of time is a stochastic quantity with a probability distribution, $P(\eta)$. Recently it has been shown that efficiency fluctuations also exhibit universal statistical properties in both classical \cite{PhysRevE.90.052145_Universal_theory_of_efficiency, Gingrich_2014Efficiency_large_deviations, PhysRevE.93.052123_eff_flu_small_machines,Verley2014_unlikely_Carnot_efficiency, PhysRevLett.114.050601Efficiency_Statistics_at_All, martinez2016_brownian_carnot_engine,PhysRevE.92.032105_eff_single_particle_engines, Proesmans_2015_effi_fluc_effusion} and quantum \cite{PhysRevB.91.115417_eff_fluc_q_thermo_ele_devices, PhysRevLett.115.040601_eff_fluct_broken_time_rev_sym, PhysRevB.92.245418_fcs_vibration} systems. In particular, for long-range autonomous machine trajectories, the distribution $P(\eta)$ concentrates on the macroscopic efficiencies, while the reversible efficiencies, become asymmetrically less probable. In addition, the Efficiency Large Deviation Function (LDF), which represents the long-run behavior, $t^{-1} \ln P(\eta)$, shows a smooth profile with only two extrema and well-defined limits for quantifying significant efficiency fluctuations. These theoretical predictions have been confirmed experimentally in Brownian engines\cite{martinez2016_brownian_carnot_engine, PhysRevX.6.041010_brownian_duet}. \par
Another fundamental aspect of quantum systems is the existence of quantum coherences. Quantum coherences  can be exploited to extract work from a single heat bath \cite{allahverdyan2011work_single, scully2003extracting_single}, reduce recombination in photocells \cite{PhysRevLett.104.207701_photocell}, and produce lasing without population inversion\cite{PhysRevLett.100.173601_lasing}. Coherences have a significant impact on the efficiency of QHEs, with some studies indicating that it can enhance efficiency beyond classical limits\cite{PhysRevA.86.043843_rahav, scully2011quantum_noise, PhysRevLett.104.207701_photocell}. Studies have focused on the nonequilibrium thermodynamics of quantum coherence, aiming to clarify the conditions under which coherence is beneficial or detrimental for work extraction from a system. The performance of quantum Otto heat engines has also been studied in the presence of coherence, with findings indicating that coherence can influence engine performance in terms of engine efficiency and power output through dynamical interference effects\cite{PhysRevA.99.062103_otto_coh}. The relationship between coherences and efficiency fluctuations in quantum heat engines can be understood through this concept of dynamical interference effects. When coherence is present along the cycle, careful tuning of the cycle parameters can exploit this interference effect, effectively making coherence act as a dynamical quantum lubricant that influences the engine's performance\cite{PhysRevA.99.062103_otto_coh}. In this work, we focus on studying the effect of noise induced coherences\cite{scully2011quantum_noise, sarmah2023learning, PhysRevA.107.052217_mjs} on efficiency fluctuations in a well established coherent quantum heat engine (QHE) prototype.  In Sec. \ref{model}, we briefly illustrate the QHE model and then present the model equations. In Sec. \ref{fcs} we show the full counting statistics of the QHE and calculate the efficiency LDF. In Sec. \ref{eff-cum}, we numerically calculate the first and second efficiency cumulants of the QHE, examining their relationship with noise induced coherences and deducing a bound between these cumulants. Finally, we conclude our work in Section \ref{conc}.

\section{Formalism}

\label{model}
\begin{figure}[!tbp]
\centering
\includegraphics[width=8.5cm]{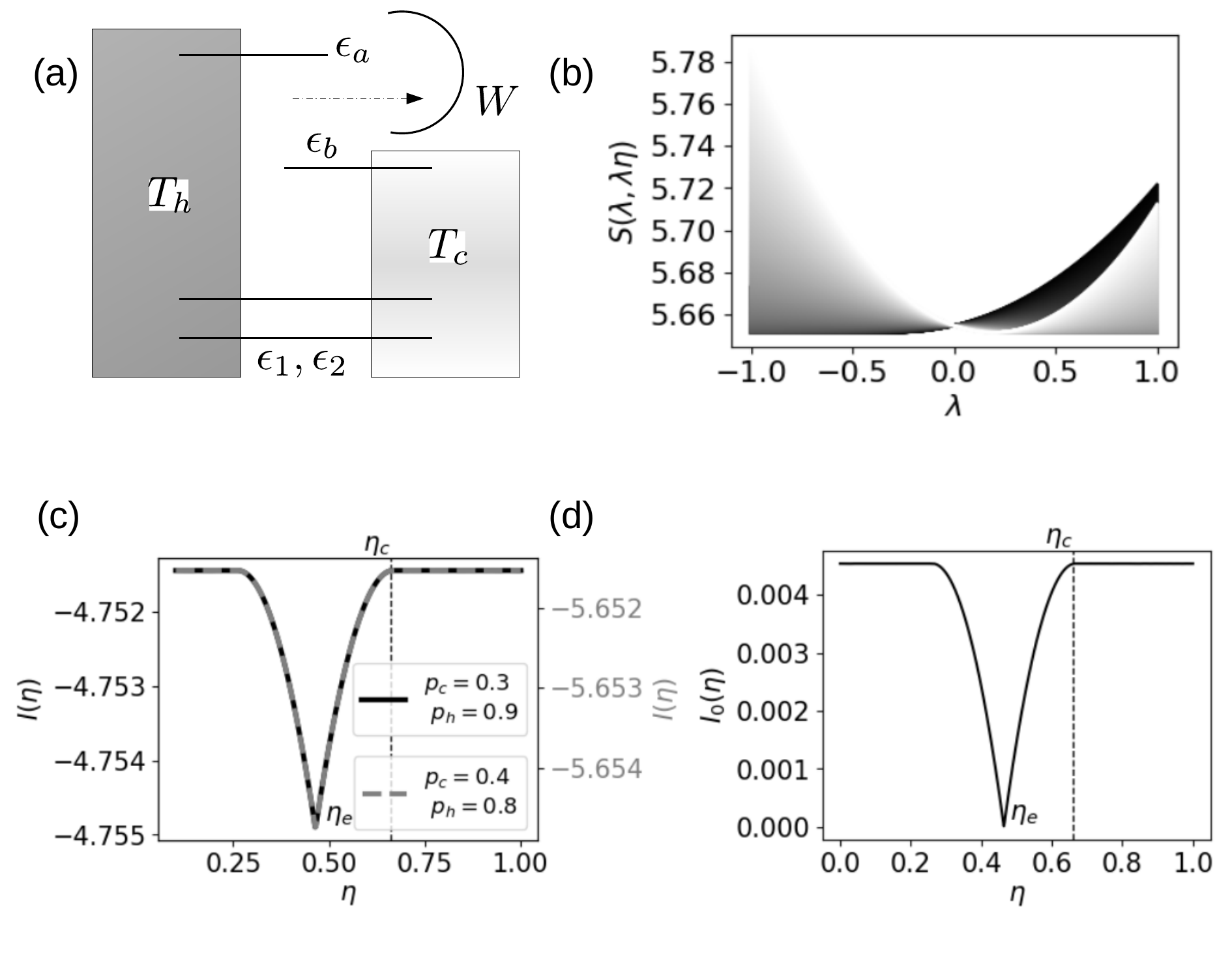}
\caption{Level scheme of the model quantum heat
engine. A pair of degenerate levels $|1\rangle$ , $|2\rangle$ is resonantly
coupled
to two excited levels $|a\rangle$ and $|b\rangle$ by two thermally populated
field modes with hot ($T_h$) and cold ($T_c$) temperatures. Levels
$|a\rangle$ and $|b\rangle$ are coupled through a nonthermal (cavity) mode of
frequency $\nu_\ell$. The emission of photons into this mode is the work
done by the QHE. The engine parameters are $T_c = 0.3,\ T_h = 0.9,\ T_l = 2, \ \epsilon_a = 1.5, \ \epsilon_1 = \epsilon_2 = 0.1, \epsilon_b = 0.5, \ g = 1,\ r = 0.7 $  in the unit of $k_B\to 1$ and $\hbar\to 1$. 
}
\label{fig1}
\end{figure}

The QHE model consists of four quantum levels coupled asymmetrically to 
two thermal reservoirs with the upper two levels coupled to a unimodal cavity as shown schematically in Fig.\ref{fig1}(a). Experimentally, similar QHEs have been realized in cold Rb and Cs atoms using magneto-optical traps  \cite{zou2017quantum,bouton2021quantum}. The quantum cavity's Hamiltonian is $\hat H_\ell= \epsilon_\ell\hat{a}_{\ell}^{\dag} \hat{a}_{\ell}$ and $\hat H_\nu=\sum_{k}\epsilon_{\nu k}\hat{a}_{\nu k}^{\dag}\hat{a} _{\nu k}$ is the Hamiltonian for the $\nu$-th reservoir. The total Hamiltonian of the four-level QHE is $\hat{H}_{T}\,=\,\sum_{m\,=\,1,2,a,b}\epsilon_{m}|m \rangle\langle m |+\hat H_\ell+\hat H_\nu+\hat{V}_{sb}+\hat V_{s\ell}$, with the system-reservoir ($\hat{V}_{sb}$) and system-cavity coupling ($\hat{V}_{s\ell}$) Hamiltonians given by,
\begin{align}
\hat V_{sb}&=\sum_{k\,\in\nu=h,c}\sum_{i\,=\,1,2}\sum_{\alpha=\,a,b}g_{ik}\hat{a}_{\nu k}|\alpha\rangle\langle i|+h.c\\
\hat V_{s\ell}&=g\hat{a}_{\ell}^\dag|b\rangle\langle a|+h.c.
\end{align}
 $\epsilon_{k}$, $\epsilon_{\ell}$ and $E_{m}$ denote the energy of the $k$th mode of the two thermal reservoirs, the unimodal cavity and system's $m$th energy level respectively. The system-reservoir coupling of the $i$th state with the $k$th mode of the reservoirs is denoted by $g_{ik}$. $\hat{a}^\dag (\hat{a})$ are the bosonic creation (annihilation) operators. 
The radiative decay originating from the transition $|a\rangle\to|b\rangle$ is the work done by the engine. The stochastic efficiency is related to this work.

\section{Full Counting Statistics}
\label{fcs}
To compute the full statistics of efficiency fluctuations, we
follow the standard two-point measurement procedure \cite{RevModPhys.81.1665_harbola_rev, PhysRevB.92.245418_bijay_prb2015}. The generating function in the Heisenberg picture
for the engine can be defined as,
\begin{align}
\label{eq-generating}
 {\cal G}(\boldsymbol{\lambda}):=\langle e^{-i\hat H_T(\boldsymbol{\lambda})t}\hat\rho_Te^{i\hat H_T(-\boldsymbol{\lambda})t}\rangle
\end{align}
where $H_T(\boldsymbol{\lambda})=\hat H_o+\hat V_{s\ell}+\hat V_{sc}(\lambda_w)+\hat V_{sh}(\lambda_q)$ with
\begin{align}
 \hat V_{sh}(\lambda_q)&=\displaystyle e^{-\lambda_q/2}\sum_{k,i=1,2}g_{ik}\hat a_{hk}|a\rangle\langle i|+H.c.,\\
 \hat V_{sc}(\lambda_w)&=e^{-\lambda_w/2}\sum_{k,i=1,2}g_{ik}\hat a_{ck}|a\rangle\langle i|+H.c.,
\end{align}
 $\lambda_q$ and $\lambda_w$ are the counting fields for tracking the heat and work cumulants respectively at the hot and cold terminals of the engine denoted by the set $\boldsymbol{\lambda}=\{\lambda_q,\lambda_q\}$. $\langle\ldots\rangle$ represents an expectation value with respect to the total density matrix $\hat\rho_T(0)$ at the initial time.   We assume that $\hat\rho_T(0)=\hat\rho_h(0)\otimes\hat\rho_c(0)\otimes\hat\rho_\ell(0)\otimes\rho(0)$, a factorized product form for the thermal bath, cavity, and system density matrix. The baths are maintained in equilibrium at $T_\nu=1/\beta_\nu,\nu =h,c,\ell$ and the states are described by the grand canonical distribution function, $\rho_\nu(0)=\exp\{-\beta_\nu\hat H_\nu\}/tr_\nu\{-\beta_\nu\hat H_\nu\}$. 
 Taking the time derivative of Eq. (\ref{eq-generating}) and tracing over only the reservoir and cavity density matrices, we can recast Eq.(\ref{eq-generating}) in the interaction picture as,
\begin{align}
 \label{eq-rho-U-dot}
 \dot{\tilde{{\cal G}}}_{}(t,\boldsymbol{\lambda})&=i
 \langle\tilde\rho_{T}(t,\boldsymbol{\lambda})\tilde V(t,\boldsymbol{\lambda})-\tilde V(t,-\boldsymbol{\lambda})\rho_{T}(t,\boldsymbol{\lambda})\rangle_{\nu,\ell}.
\end{align}

The interaction picture is defined as $\tilde O(t):=\exp(i \hat H_o t)\hat O\exp(-i \hat H_ot)$, with $\hat H_o$ being the bare or noninteracting  Hamiltonian of the engine and $\hat V$ is the total coupling Hamiltonian. 
 By formally integrating Eq.(\ref{eq-rho-U-dot}) we keep terms up to second order in the perturbation of the coupling, $\hat{V}= \hat{V}_{sb}+\hat{V}_{s\ell}$.We then substitute the operators $\hat V_{sb}$ and $\hat V_{s\ell}$ in Eq.(\ref{eq-rho-U-dot}) and invoke the  Born-Markov and secular approximation by assuming a flat and constant density of states for each coupling (wide band approximation). The time evolution of the reduced density matrix in the Schrodinger picture now reads (using the short form of the operators $|i\rangle\langle j|= B_{ij}$),
\begin{align}
\label{eq-pre-qme}
&\dot {\cal G}(t,\boldsymbol{\lambda})=-[\hat{H}_o,\rho(t,\boldsymbol{\lambda})]-\pi\Omega \nonumber\\&\times\displaystyle\sum_{i,j=1,2}\Big{[}g_{ih}g^*_{jh}\tilde n_h(B^\dag_{ia}B_{ja}\rho(t,\boldsymbol{\lambda})
-e^{\lambda_q}B^\dag_{ia}\rho(t,\boldsymbol{\lambda})B_{ja})\nonumber\\&-g_{jh}g^*_{ih}\tilde n_h(e^{\lambda_q}B^{}_{ia}\rho(t,\boldsymbol{\lambda})B_{ja}^\dag-\rho(t,\boldsymbol{\lambda})B^\dag_{ia}B_{ja})\nonumber\\
&+g_{ih}g^*_{jh}n_h(B^\dag_{ia}B_{ja}\rho(t,\boldsymbol{\lambda})-e^{-\lambda_q}B^\dag_{ia}\rho(t,\boldsymbol{\lambda})B_{ja})\nonumber\\
&-g_{jh}g^*_{ih}n_h(e^{-\lambda_q}B^{}_{ia}\rho(t,\boldsymbol{\lambda})B_{ja}^\dag-\rho(t,\boldsymbol{\lambda})B_{ia}B^\dag_{ja})\nonumber\\
&+g_{ic}g^*_{jc}\tilde n_c(B^\dag_{ib}B_{jb}\rho(t,\boldsymbol{\lambda})-e^{\lambda_w}B^\dag_{ia}\rho(t,\boldsymbol{\lambda})B_{ja})\nonumber\\
&-g_{jc}g^*_{ic}\tilde n_c(e^{\lambda_w}B^{}_{ib}\rho(t,\boldsymbol{\lambda})B_{jb}^\dag-\rho(t,\boldsymbol{\lambda})B^\dag_{ib}B_{jb})\nonumber\\
&+g_{ic}g^*_{jc}n_c(B^\dag_{ib}B_{jb}\rho(t,\boldsymbol{\lambda})-e^{-\lambda_w}B^\dag_{ib}\rho(t,\boldsymbol{\lambda})B_{jb})\nonumber\\
&-g_{jc}g^*_{ic}n_c(e^{-\lambda_w}B^{}_{ib}\rho(t,\boldsymbol{\lambda})B_{jb}^\dag-\rho(t,\boldsymbol{\lambda})B_{ib}B^\dag_{jb})\Big{]}\nonumber\\
&-\pi \Omega g^2\nonumber\\ &\!\times\!\Big{[}\tilde n_{\ell}\big{[}B^\dag_{ba}B_{ba}\rho(t,\boldsymbol{\lambda})-2B^\dag_{ba}\rho(t,\boldsymbol{\lambda})B_{ba}\!+\!\rho(t,\boldsymbol{\lambda})B^\dag_{ba}B_{ba}\big{]}\nonumber\\&
\!-\!n_\ell\big{[}B_{ba}B^\dag_{ba}\rho(t,\boldsymbol{\lambda})\!-\!2B_{ba}\rho(t,\boldsymbol{\lambda})B^\dag_{ba}\!+\!\rho(t,\boldsymbol{\lambda})B_{ba}B^\dag_{ba}\big{]}\Big{]}
\end{align}
Using the definition of the density matrix elements, $\rho_{mn}:=\langle m|\rho(t)|n\rangle, m = 1,2,a,b$, in Eq. (\ref{eq-pre-qme}), we obtain four populations ($\rho_{mm},m=1,2,a,b)$ and a real part of the coherence $\rho_{12}$ that describe the time evolution of the reduced system density matrix. The double excitation operators such as $B_{ib}, B_{jb}$, etc also do not generate additional terms and hence these do not contribute to the dynamics. 
Under the symmetric and real coupling regime ($g_{i\nu}=g_{j\nu'}=2r/\pi\Omega, \nu,\nu'\in h,c$ and $\pi \Omega g^2 =r$ ), the generating function can be effectively written as, $\dot{\cal G}(\boldsymbol{\lambda}) = \langle\mathcal{L}(\boldsymbol{\lambda})|\rho(\boldsymbol{\lambda})\rangle$ with the effective twisted generator ${\cal L}(\boldsymbol{\lambda})=r\times$
\begin{small}
\begin{equation}
\label{Louv-eq}
\begin{pmatrix}
-\displaystyle n&0&\tilde n_he^{\lambda_q}&\tilde n_ce^{\lambda_W}&-y\\[2mm]
0&-\displaystyle  n&\tilde n_he^{\lambda_q}&\tilde n_ce^{\lambda_W}&-y\\[2mm]
n_he^{-\lambda_q}&n_he^{-\lambda_q}& -\displaystyle\frac{
\tilde G}{r}&\displaystyle\frac{g^2 n_\ell}{r} &2p_hn_he^{-\lambda_q}\\[2mm]
 n_c e^{-\lambda_W}& n_ce^{-\lambda_W}&\displaystyle\frac{g^2_{}\tilde n_\ell}{r}&-\displaystyle\frac{G}{r}&2p_cn_ce^{-\lambda_W}\\[2mm]
-y/2&-y/2&p_h\tilde n_h e^{\lambda_q}&p_c\tilde n_ce^{\lambda_W}&-n
\end{pmatrix}\\[2mm]
\end{equation}
\end{small}
$\tilde G=(g^2\tilde n_\ell+2r\tilde n_c), G=(g^2n_\ell+2r\tilde n_c), n=n_c+n_h, y =n_cp_c+n_hp_h$ and $\langle\rho(\boldsymbol{\lambda})| =\{\rho_{11}(\boldsymbol{\lambda}),\rho_{22}(\boldsymbol{\lambda}),\rho_{aa}(\boldsymbol{\lambda}),\rho_{bb}(\boldsymbol{\lambda}),\rho_{12}(\boldsymbol{\lambda})\}$.
The terms $
 n_\ell, n_c $ and $ n_h$
represent the Bose-Einstein thermal occupation factors at the energies $\epsilon_a-\epsilon_b, \epsilon_b-\epsilon_1$ and $\epsilon_a-\epsilon_1$ respectively with $\tilde n_\nu=1+n_\nu,\nu =h,c,\ell$. The two dimensionless parameters, 
$p_\nu,\nu=h,c$ are a measure of the strength of
coherences \cite{PhysRevA.88.013842_hpg01, doi:10.1073/pnas.1212666110Photosynthetic_reaction_center_as_a_qhe, PhysRevB.91.115417_eff_fluc_q_thermo_ele_devices, PhysRevA.86.043843_Rahav_Reducued_DM_1, Harbola_2012Reducued_DM_2, PhysRevE.99.022104_giri_ml, sarmah2023learning} whose values are dictated by the angles of the relative orientation of the transition dipoles. The 
subscripts $c$ and $h$ are used to keep track of 
contributions coming from
couplings to the cold and hot reservoirs, respectively. These are referred to as the noise induced coherences. The physical significance and parametrization technique of introducing these two parameters have been discussed and used a lot in earlier literature \cite{PhysRevA.88.013842_hpg01, doi:10.1073/pnas.1212666110Photosynthetic_reaction_center_as_a_qhe, PhysRevB.91.115417_eff_fluc_q_thermo_ele_devices, PhysRevA.86.043843_Rahav_Reducued_DM_1, Harbola_2012Reducued_DM_2, PhysRevE.99.022104_giri_ml, sarmah2023learning}.

In this formalism, the stochastic heat
$Q_h$ is simply the difference between the thermal energies at the levels $|1\rangle$ and $|a\rangle$,  $\epsilon_a-\epsilon_1$. The stochastic work ($W$), however, is not the energy difference between the upper two states $|a\rangle$ and $|b\rangle$ and needs to be identified since the latter are coupled to a quantum cavity where dissipation exists due to spontaneous emission. To estimate the actual quantum work, we first assume that $\lambda_W=0$. In this case, a known identity, reminiscent of the  Gallavoti-Cohen linear-shift symmetry leading to a steady-state fluctuation theorem is known to hold\cite{PhysRevA.86.043843_rahav}. Mathematically, ${\cal U}^{-1}{\cal L}(\lambda_q){\cal U}={\cal L}^T(-\lambda_q+ie^{\cal F})$ with ${\cal U} =diag\{r,r,r,\frac{n_cn_\ell}{\tilde n_c\tilde n_l}, r\frac{n_c}{\tilde n_c},r/2\}$. ${\cal F}$ is the thermodynamic force that drives the system out of equilibrium with,
\begin{align}
 \ln{\cal F}=\displaystyle\frac{Q_h}{k_BT_h}-\frac{Q_c}{k_BT_c}+\ln\frac{n_\ell}{\tilde n_\ell}, 
\end{align}
where $Q_h=\epsilon_a-\epsilon_1, Q_c= \epsilon_b-\epsilon_1$. The first two terms are the entropies at the hot and cold terminals while the last term is the dissipation into the quantum cavity and needs to be accounted for in the thermal gradient $Q_h-Q_c$ during the heat-to-work conversion process\cite{PhysRevA.88.013842_hpg01, PhysRevE.99.022104_2019_hpg_ml}. Hence, the useful stochastic work can be defined as,
\begin{align}
\label{eq-W}
 W&=\epsilon_a-\epsilon_b+k_BT_c\ln\frac{\tilde n_\ell}{n_\ell}\\
 &=(\epsilon_a-\epsilon_1)\eta_C-k_BT_c\ln{\cal F}
\end{align}
   with $\eta_C$ being the Carnot efficiency. To account for the efficiency fluctuations we can transform the individual counting fields into an efficiency tracking field, $\lambda$ by substituting $\lambda_q=\eta \lambda$ and $\lambda_W=\lambda$ \cite{PhysRevB.92.245418_bijay_prb2015, Verley2014_unlikely_Carnot_efficiency} so that ${\cal L}(\boldsymbol{\lambda})= {\cal L}(\lambda,\eta\lambda)$ with $\eta$ being the stochastic efficiency. In the long time limit, we  define a scaled cumulant generating function,
\begin{align}
 S(\lambda,\eta\lambda)=\displaystyle\lim_{t\to\infty}\frac{1}{t}\ln \langle\rho(\lambda,\eta\lambda,t)\rangle.
\end{align}

The stochastic efficiency cumulant generating function, $S(\lambda,\eta\lambda)$ can be evaluated from Eq. (\ref{Louv-eq}). In the steady state, when $\boldsymbol{\lambda}= 0$, a zero eigenvalue is obtained from the RHS of Eq.(\ref{Louv-eq}). This zero-eigenvalue corresponds to the cumulant generating function, $S(\lambda,\eta\lambda)$ \cite{esposito2009nonequilibrium}.
For different $\eta$ values, we show the variation of $S(\lambda,\eta\lambda)$ for fixed coherences in Fig. (\ref{fig1}b) by using the definition of $W$ from Eq. (\ref{eq-W}) along with $Q_h=\epsilon_a-\epsilon_1$.  It is piecewise smooth and concave downwards for all values of $0\le \eta\le 1$ with a single minima as a function of the efficiency tracking field $\lambda$. 
From the cumulant generating function,
$S(\lambda,\eta\lambda)$, we can define the large-deviation function via the extremization\cite{Verley2014_unlikely_Carnot_efficiency},
\begin{align}
 I(\eta)=-\displaystyle\min_{\lambda}S(\lambda,\eta\lambda).
\end{align}
We numerically minimize $S(\lambda,\eta\lambda)$ and evaluate the large deviation function for the efficiency, $I(\eta)$ and show it in Fig. (\ref{fig1}c) for two different combinations of the coherence measure parameters $p_c$ and $p_h$. For both the combinations, $I(\eta)$ exhibits a single minima at $\eta=\eta_e=W/Q_h$ (0.46 in the figure, which corresponds to the most likely efficiency) with the expected maxima at the Carnot efficiency, $\eta_C$ (0.66 in the figure, the least likely efficiency).  Apart from a change in magnitude, the nature of the $I(\eta)$  curve doesn't change for different values of coherences.  We also evaluate the large deviation function in the absence of coherences ($p_c=p_h=0,$ so $\rho_{12}=0$), which we denote as $I_o(\eta)$ which is also seen to yield the macroscopic quantum efficiency, $\eta_e$ as the most likely efficiency in Fig. (\ref{fig1}d). 


\section{Efficiency Cumulants}
\label{eff-cum}
\begin{figure}[!tbp]
\centering
\includegraphics[width=8.5cm]{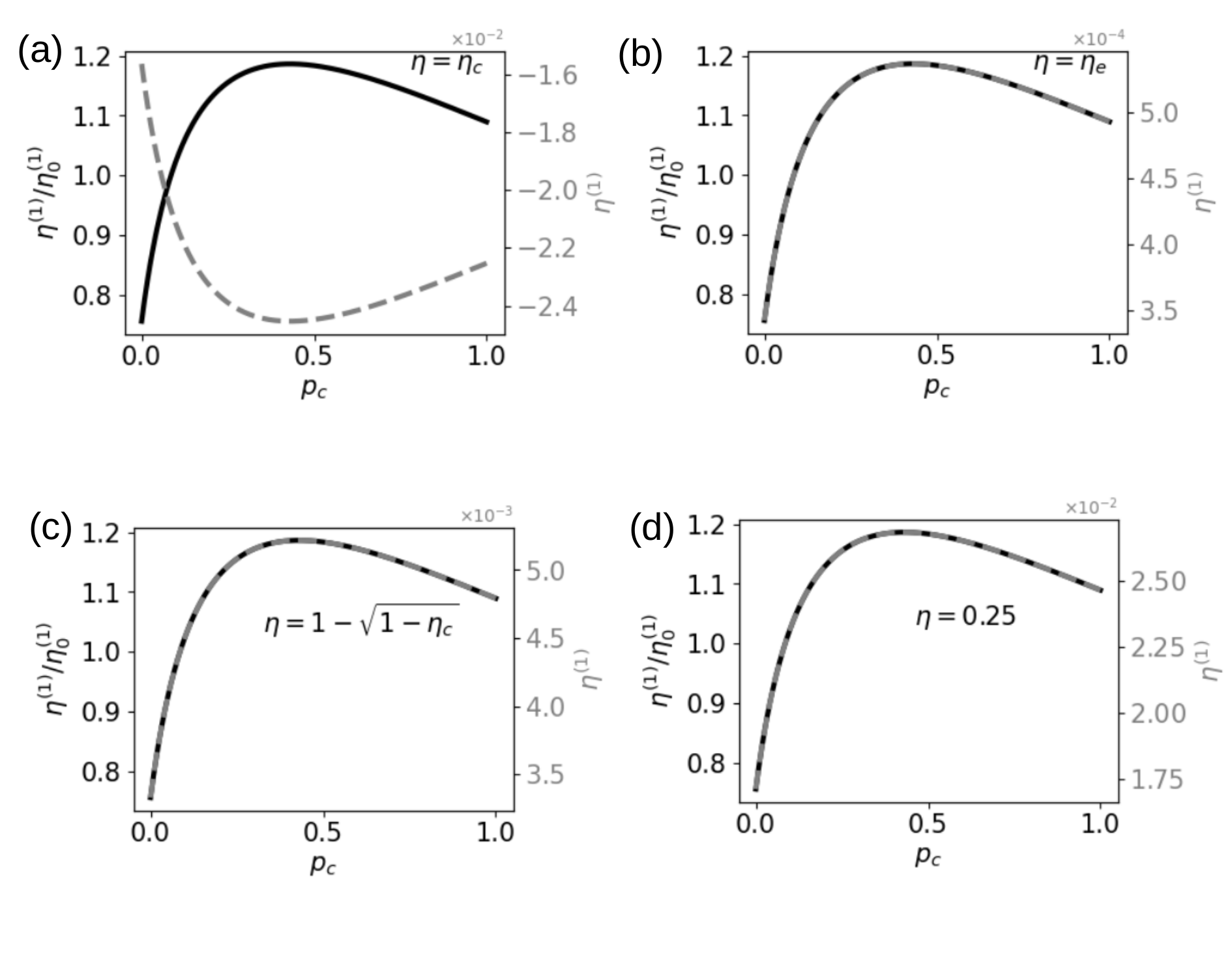}
\caption{The first cumulant $\eta^{(1)}$ as a function of the cold coherence parameter, $p_c$ for fixed $p_h$ at four different $\eta$ values, (a)$\eta=\eta_C$, (b) $\eta=\eta_e$,  the macroscopic quantum efficiency (c) at the efficiency at maximum power and (d) at a random value of the efficiency. Note that the absolute value of the first cumulant is negative when $\eta=\eta_C$. For other efficiencies, it is positive. Note the constancy of the ratio $\eta^{(1)}/\eta^{(1)}_o$ at all efficiencies. The engine parameters are fixed through out the manuscript at $\epsilon_1=\epsilon_2=0.1, \epsilon_b=0.4, \epsilon_a = 1.5, g = 1, r = 0.7 $  and $\tau = 0.5$ in the unit of $k_B\to 1$ and $\hbar\to 1$. 
}
\label{fig2}
\end{figure}
\begin{figure}[!tbp]
\centering
\includegraphics[width=8.5cm]{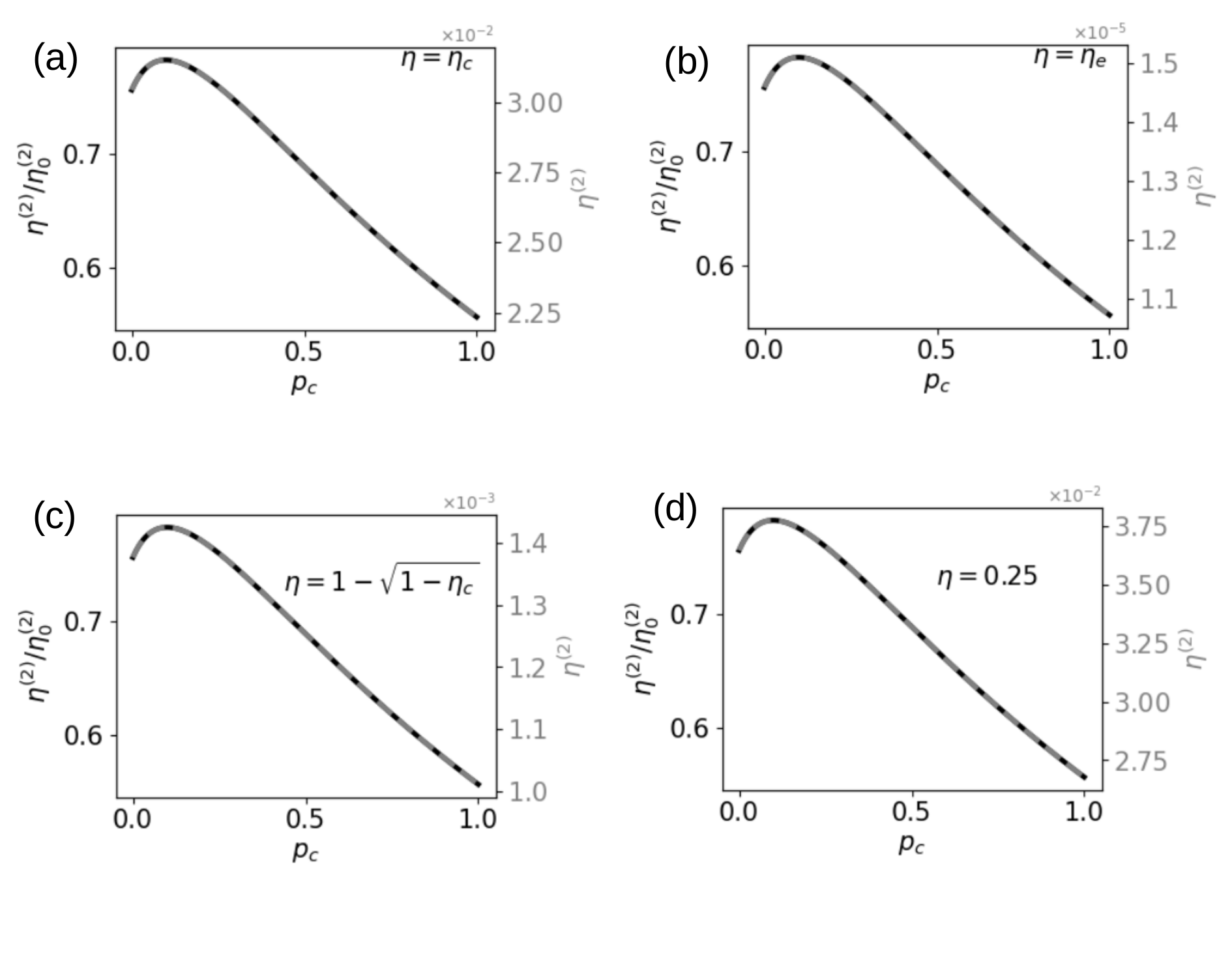}
\caption{The second cumulant $\eta^{(2)}$ as a function of the cold coherence parameter, $p_c$ for fixed $p_h$ at four different $\eta$ values, (a)$\eta=\eta_C$, (b) $\eta=\eta_e$,  the macroscopic quantum efficiency (c) at the efficiency at maximum power and (d) at a random value of the efficiency. Note that the absolute value of the second cumulant is always positive even at $\eta=\eta_C$, unlike the first cumulant. Note the constancy of the ratio $\eta^{(2)}/\eta^{(2)}_o$ at all efficiencies, although the individual $\eta^{(2)}$ values are significantly different.
}
\label{fig3}
\end{figure}
\begin{figure}[!b]
\centering
\includegraphics[width=8.5cm]{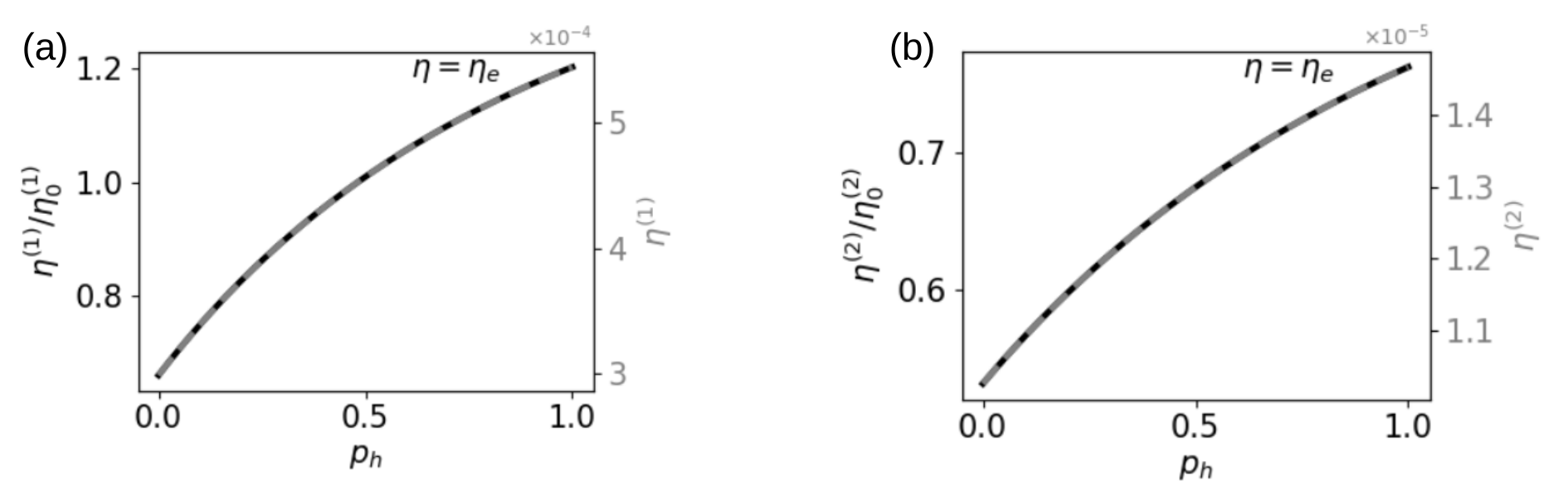}
\caption{The positive first and second cumulants as a function of the hot coherence parameter, $p_h$ for fixed $p_c$ at the macroscopic efficiency $\eta=\eta_e$.  Note the constancy of the ratio $\eta^{(i)}/\eta^{(i)}_o$. }
\label{fig4}
\end{figure}
The $n$-th cumulant of the stochastic quantum efficiency can be obtained as
\begin{align}
\label{cum-eq}
 \eta^{(n)}&=\partial_\lambda^n S(\lambda,\eta\lambda)|_{\lambda=0}.
\end{align}
We denote the cumulants in the absence of coherences ($p_\nu=0$) as $ \eta^{(n)}_o$. When $\eta^{(n)}>(<)\eta^{(n)}_o$, the coherences increases (decreases) the value of the $n$-th cumulant in comparison to the classical case. 
The variation of  the first quantum efficiency cumulant $\eta^{(1)}/\eta^{(1)}_o$ as a function of the cold coherence parameter, $p_c$ is shown in Fig. (\ref{fig2}a-d) for a fixed value of $p_h$ and four different $\eta$ values, a)$\eta = \eta_c$, the Carnot efficiency, b)$\eta = \eta_e = W/Q_h$, the most likely macroscopic efficiency, c) $\eta = 1-\sqrt{1-\eta_c}$, the efficiency at maximum power and d) $\eta = 0.25$, a randomly selected efficiency. The variation of  $\eta^{(1)}$ is also shown in the same figure with a different axis. When $\eta=\eta_C$, $\eta^{(1)}$ is negative as shown by RHS axes of Fig. (\ref{fig2}a). The ratio, $\eta^{(1)}/\eta^{(1)}_o$, is however positive since $\eta^{(1)}_o$ is negative. $\eta^{(1)}/\eta^{(1)}_o$ contain regions both greater and less than unity as a function of $p_c$ highlighting that the noise-induced coherence can increase as well as decrease the cumulant's value in comparison to the classical case. The ratio has an optimal  value similar to what has been previously reported for the flux (first moment of particle transport) as a function of thermally induced coherence. $\eta^{(1)}$ is however positive when the efficiency cumulant is evaluated at the macroscopic quantum efficiency $\eta_e$, Fig. (\ref{fig2}b). It is also positive at the efficiency value when the power is maximum, i.e the Curzon-Ahlborn efficiency, $1-\sqrt{1-\eta_C}$ (Fig. (\ref{fig2}c)) as well as at a randomly chosen efficiency (Fig. (\ref{fig2}d). Note that the shape of all the curves of the first cumulant, $\eta^{(1)}$ at the four different values of $\eta$ are identical with the exception when $\eta=\eta_C$ as the curve is rotated because of negative $\eta^{(1)}$ values.
Hence there is an existence of the following identity,
\begin{align}
 \displaystyle\frac{\eta^{(i)}}{\eta^{(i)}_o}=K^{(i)}, \forall \eta \in (-\infty,\infty),
\end{align}
where $K^{(i)}$ is a nonlinear function of the coherence parameters but independent of the stochastic efficiency. 
 Further, the ratio $\eta^{(1)}/\eta^{(1)}_o$ is constant and identical even though the individual numerator and the denominator are different. We found this to be true for all values of $\eta$ evaluated from the numerical cumulant generating function, $S(\lambda,\eta\lambda)$ in Fig. (\ref{fig1}b). The constancy in the ratio between the first cumulants of the quantum efficiency in the presence and absence of coherences is also present in the second cumulant for all $\eta$ values as seen from Fig. (\ref{fig3}a-d). The magnitude of $\eta^{(2)}$ is different and always positive in the interval $0\le p_c\le 1$ and $\eta^{(2)}$ is a different constant at all the $\eta$ values. The ratio, $\eta^{(2)}/\eta^{(2)}_o$, is however bound below unity indicating that the cold bath-induced coherence decreases the second quantum efficiency cumulant in comparison to the classical value. The behavior of the cumulant is similar to that of the first cumulant where an increase followed by a decrease is observed. However, the optimal value of $p_c$ where the second cumulant is maximum occurs at a lower value of $p_c$ than what was seen for the first cumulant. The $\eta-$ independency is also observed when the ratio between the quantum and classical cumulants is evaluated in the hot bath-induced coherence interval $0\le p_h\le 1$. Both $\eta^{(i)}$ and $\eta^{(i)}_o$ are different at $\eta=\eta_e$, but the ratio remains constant. The independency of the ratio $\eta^{(i)}/\eta^{(i)}_o$ can be analytically understood near equilibrium ($\ln{\cal F}\approx 0)$ . In this case, we expand the exponential dependence of the tracking fields in the twisted generator of Eq. (\ref{Louv-eq}) \cite{PhysRevA.88.013842_hpg01, Harbola_2012Reducued_DM_2} and keep terms upto first order, i.e,  $\exp(Q_h\eta\lambda)\approx 1+Q_h\eta\lambda$ and $\exp(W\lambda)$ $\approx$ $ 1+W\lambda$) so that ${\cal L}(\lambda,\eta\lambda)\approx {\cal L} (0,0)+{\cal L}^{(1)}(\lambda,\eta\lambda)$ with ${\cal L}^{(1)}(\lambda,\eta\lambda)=r\times$ 
\begin{small}
\begin{equation}
\label{Louv1-eq}
\begin{pmatrix}
\displaystyle 0&0&\tilde n_h Q_h\eta\lambda&\tilde n_c W\lambda&0\\[2mm]
0&0&\tilde n_h Q_h\eta\lambda&\tilde n_cW\lambda&0\\[2mm]
-n_hQ_h\eta\lambda&-n_hQ_h\eta\lambda& 0&0 &-2p_hn_hQ_h\eta\lambda\\[2mm]
 -n_c W\lambda& -n_cW\lambda&0&0&-2p_cn_cW\lambda\\[2mm]
0&0&p_h\tilde n_h Q_h\eta\lambda&p_c\tilde n_cW\lambda&0
\end{pmatrix}\\[2mm]
\end{equation}
\end{small}
${\cal L}(0,0)$ has a single positive eigenvalue with four negative eigenvalues. The positive eigenvalue $S^{(0)}$ is independent of $\eta$ but is not analytically identifiable while the eigenvalue of ${\cal L}^{(1)}(\lambda,\eta\lambda)$ is given by
\begin{align}
 S^{(1)}(\lambda,\eta\lambda)&=
 \eta\lambda Q_hr \displaystyle\sqrt{t_1+\sqrt{t_2+t_1^2}}\\
   t_1&=-\tilde n_c^2 \left(p_c^2+1\right) -n_h^{} \tilde{n}_h^{} \left(p_h^2+1\right)\\
   t_2&=4\tilde n_c^2 n_h^{} \tilde{n}_h^{}(p_c-p_h)^2. 
\end{align}
Note the non-Gaussian nature of the function as opposed to the Gaussian type generating functions used else where\cite{Verley2014_unlikely_Carnot_efficiency, PhysRevB.92.245418_bijay_prb2015}. When $t_1<0$, $t_1^2<t_2+t_1^2$ since $t_2>0$. The efficiency cumulant generating function can hence be approximated as $S(\lambda,\eta\lambda)\approx S^{(0)}+S^{(1)}(\lambda,\eta\lambda)$. Thus, near equilibrium, the first efficiency cumulant is simply
\begin{align}
\label{eq-cum1}
 \eta^{(1)}&=\displaystyle\partial_\lambda\{S^{(0)}+ S^{(1)}(\lambda,\eta\lambda)\}\big{|}_{\lambda\to 0}\\
 &=Q_hr\eta  \displaystyle\sqrt{t_1+\sqrt{t_2+t_1^2}}
 \end{align}
 which is linear in $\eta$. Setting $p_c=p_h=0$, we can easily obtain $\eta^{(1)}_o$ which is also the linear in $\eta$.  One can easily predict the $\eta-$ independency of the second cumulants' ratio. To evaluate the second cumulant, one needs to take the second order terms in the expansion of ${\cal L}(\lambda,\eta\lambda)$ and take a second derivative of $S(\lambda,\eta\lambda)$. The second derivative shall kill the zeroth and first-order terms and a $\eta^2$ dependence shall be observed in both $\eta^{(2)}$ and $\eta^{(2)}_o$ rendering the ratio to be independent of $\eta$. Without loss of generality, this observation of $\eta$-independency can be extended to all $\eta^{(i)}$ and $\eta^{(i)}_o$.
 Therefore $\eta^{(1)}/\eta^{(1)}_o$ is always independent of the fluctuating efficiency $\eta$. The $\eta$-independency is also valid when $\eta^{(i)}/\eta^{(i)}_o$ in  the hot coherence interval $0\le p_h\le 1$ as shown in Fig. (\ref{fig4}). The only difference is the lack of optimization since the cumulant ratio nonlinearly increases and is maximum at $p_h=1$. The second efficiency cumulant-ratio is bounded below unity as can be seen in Fig. (\ref{fig3}) for the cold coherence interval $0\le p_c\le 1$. It is also true for the coherence interval $0\le p_h \le 1$ as seen in Fig. (\ref{fig4}b).

\begin{figure}[!tbp]
\centering
\includegraphics[width=8.5cm]{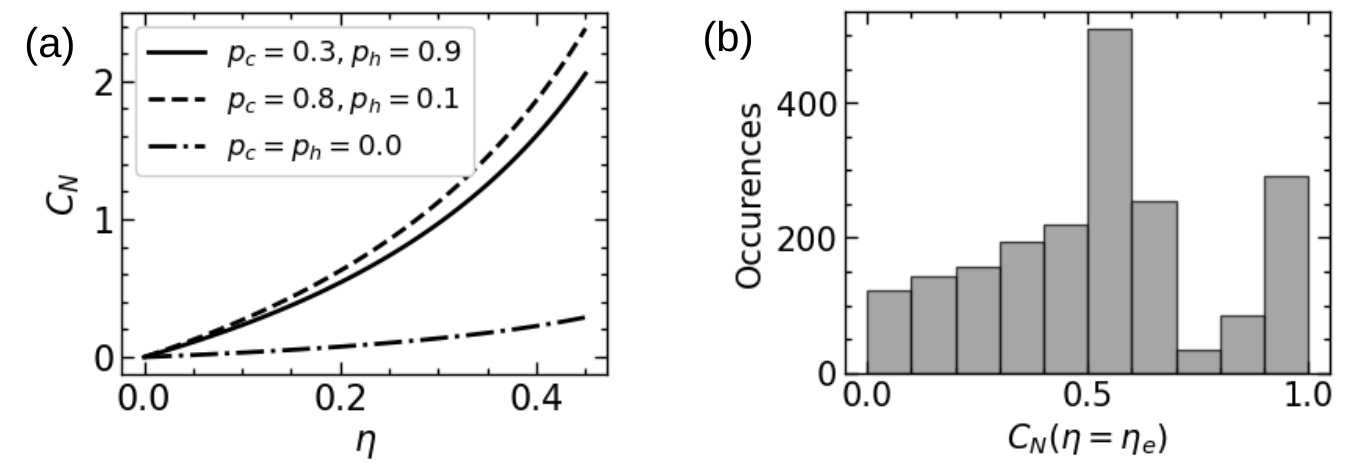}
\caption{(a) Constancy, $C_N$, as a function of $\eta$ for three different values of $p_c, \ p_h$. (b) Histogram of $C_N$  values at $\eta = \eta_e$. 
Parameters used for data generation: $0\leq p_c\leq1,\ , 0\leq p_h\leq1, 0.3\leq T_c\leq0.9,\ 0.9\leq T_h\leq 1.6,\ 2\leq T_l\leq5 $ in the unit of $k_B\to 1$ and $\hbar\to 1$. }
\label{fig-krzysztof}
\end{figure}
 \begin{figure}[!tbp]
\centering
\includegraphics[width=8.5cm]{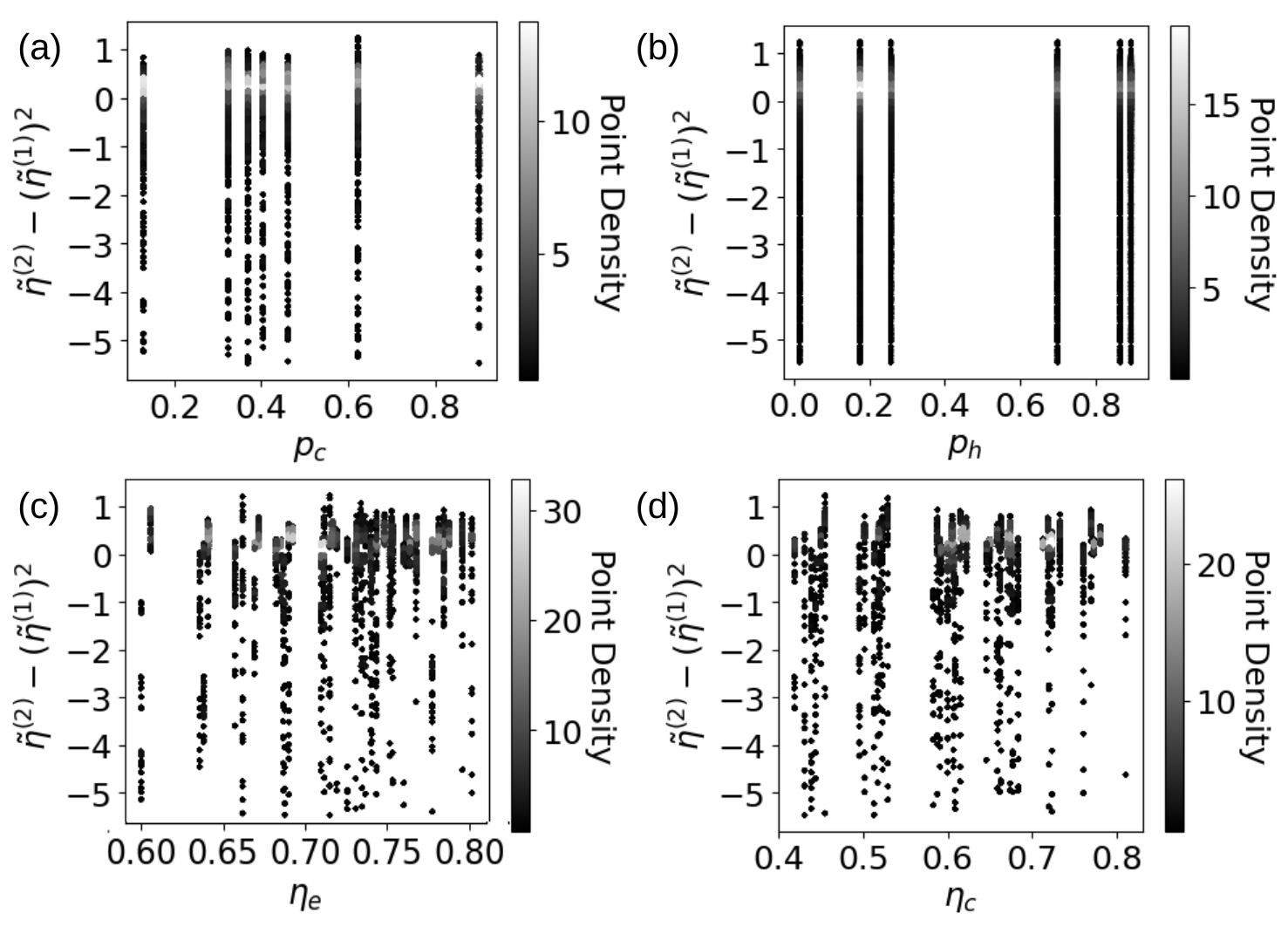}
\caption{ Values of $\tilde{\eta}^{(2)}-(\tilde\eta^{(1)})^2$ for randomly generated engine parameters $p_c$, $p_h$, $t_c$, $t_h$,and $t_l$ with (a) varying $p_c$, (b) varying $p_h$, (c) varying $\eta_e$, (d) varying $\eta_c$. For all cases Eq.(\ref{eq-saryal}) does not hold.
Parameters used for data generation: $0\leq p_c\leq1,\ , 0\leq p_h\leq1, 0.3\leq T_c\leq0.7,\ 0.2+T_c\leq T_h\leq 0.5+T_c,\ 2\leq T_l\leq7 $  in the unit of $k_B\to 1$ and $\hbar\to 1$.}
\label{fig-bound_bj}
\end{figure}

\begin{figure}[!tbp]
\centering
\includegraphics[width=8.5cm]{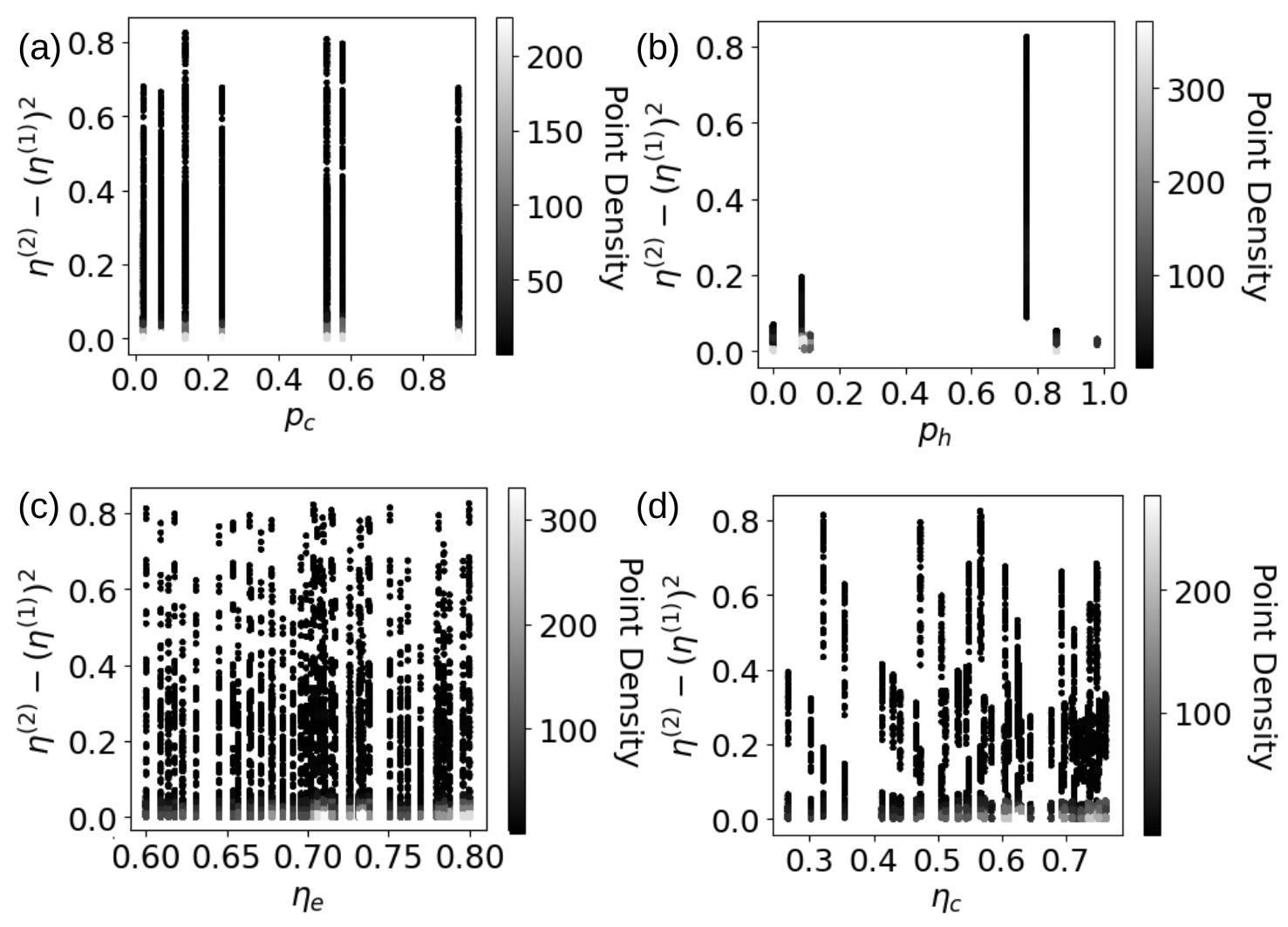}
\caption{Difference between the second and the square of the first cumulant for random values of engine parameters $p_c$, $p_h$, $T_c$, $T_h$, and $T_l$ with (a) varying $p_c$, (b) varying $p_h$, (c) varying $\eta_e$, (d) varying $\eta_c$. Note that the distributions are always positive highlighting the validity of Eq.(\ref{eq-ineq}).
Parameters used for data generation: $0\leq p_c\leq1,\ , 0\leq p_h\leq1, 0.3\leq T_c\leq0.7,\ 0.2+T_c\leq t_h\leq 0.5+T_c,\ 2\leq T_l\leq7 $  in the unit of $k_B\to 1$ and $\hbar\to 1$.}
\label{fig-bound_2}
\end{figure}

\begin{figure}[!tbp]
\centering
\includegraphics[width=8.5cm]{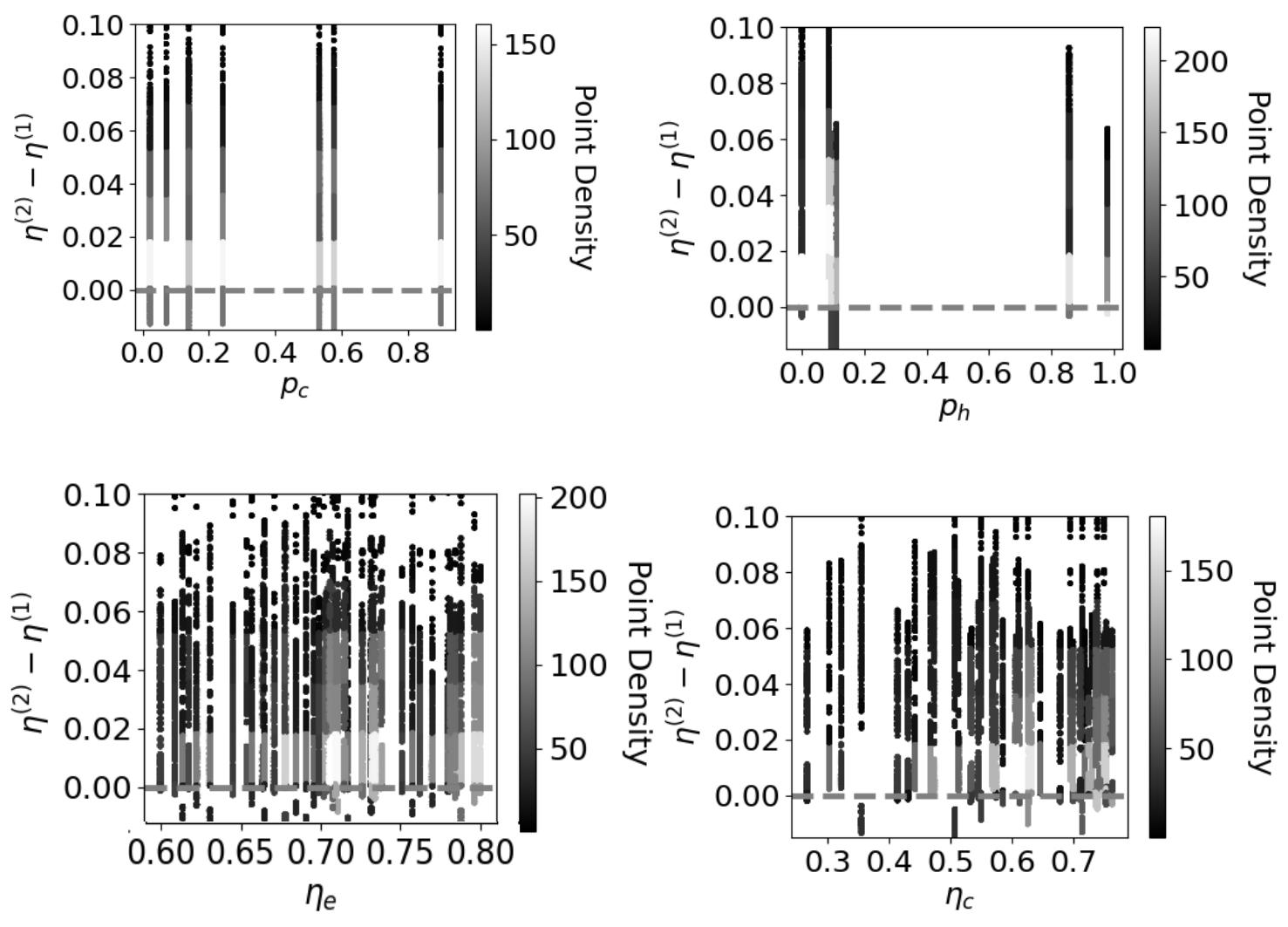}
\caption{Difference between the second and the first cumulants for random values of engine parameters $p_c$, $p_h$, $T_c$, $T_h$, and $T_l$ with (a) varying $p_c$, (b) varying $p_h$, (c) varying $\eta_e$, (d) varying $\eta_c$. Note that the distributions have both positive and negative values.
Parameters used for data generation: $0\leq p_c\leq1,\ , 0\leq p_h\leq1, 0.3\leq T_c\leq0.7,\ 0.2+T_c\leq T_h\leq 0.5+T_c,\ 2\leq T_l\leq7 $   in the unit of $k_B\to 1$ and $\hbar\to 1$.}
\label{fig-bound}
\end{figure}

It has been proposed in cyclic thermal machines that for specific working regimes the second cumulants can be suppressed by tuning coherences\cite{PhysRevB.98.085425_krytz} thereby exhibiting bounds between these. Such suppression exists to ensure the validity of the trade-off between power and efficiency. The quantification of the suppression is usually done by evaluating the particle or power fluctuations and relating it to the energy gradient (work) so that there exists a constancy in the engine. The normalized constancy is given by:
\begin{align}
    C_N = \frac{k_B T_c j^{(1)}}{j^{(2)}W} \frac{2 \eta}{\eta_c - \eta},
\end{align}
where $T_c$ is the temperature of the cold reservoir. $j^{(i)}$ is the $i-$th particle cumulant of the engine at the cold terminal which can be obtained by setting $\lambda_q = 0$, $\lambda_W = \lambda$ in Eq.(\ref{Louv-eq}) and following the standard FCS formalism. $\eta$ ($W$ ) is the nonstochastic efficiency (work )of the engine. In a standard Markovian scenario, $C_N\leq1$\cite{PhysRevB.98.085425_krytz, PhysRevLett.120.190602_seifert} when $\eta<<\eta_c$. By tuning coherences, $C_N>1$ was achieved in thermoelectric systems when the work was smaller than the thermal gradient ($T_h-T_c$). We evaluate $C_N$ using the definition of $W$ from Eq. (\ref{eq-W}) and treating $\eta$ as a stochastic variable for two different values of the coherence parameters as well as in the absence of coherences, as shown in Fig.(\ref{fig-krzysztof}(a)). From the graph, $C_N>1$ is observed as the stochastic efficiency increases. Further, in the absence of coherences, $C_N<1$ is observed (lowest curve in Fig.(\ref{fig-krzysztof}(a)). So the stochastic nature of the efficiency can be exploited only in the presence of coherences to observe $C_N>1$. We next show the numerically evaluated $C_N$ values in Fig(\ref{fig-krzysztof}(b)) for a wide range of engine parameters including coherence values at the most likely value of efficiency, $\eta = \eta_e$. In this case we do not observe $C_N>1$. Thus the constancy is robust when the engine is working at its most likely efficiency even in the regime where $W<<(T_h-T_c)$ and in the presence of coherences. Treating the efficiency as a stochastic parameter violates the constancy bound by dint of noise suppression only when aided by coherences. 
 
 To study other proposed bounds on the efficiency cumulants, we refer to a recent work on Otto engines\cite{PhysRevE.103.L060103_saryal}, where it was shown that the ratio between the second work cumulant ($W^{(2)}$) and the second heat cumulant ($Q_h^{(2)}$) is bounded below by the square of the ratio between the first work cumulant ($W^{(1)}$) and the first heat cumulant ($Q_h^{(1)}$). The fluctuating efficiency cumulants were quantified by defining the quantity, $ \tilde\eta^{(i)}=W^{(i)}/Q_{h}^{(i)}$, which is different from the actual stochastic efficiency cumulants, $\eta^{(i)}$, evaluated in this work.  An inequality of the following type was proposed\cite{PhysRevE.103.L060103_saryal},
 \begin{equation}
 \label{eq-saryal}
     \tilde\eta^{(2)} - (\tilde\eta^{(1)})^2 > 0. 
\end{equation}
 In our case the work (heat) cumulants, $W^{(i)}$, ($Q_h^{(i)}$) are obtained by setting $\lambda_q=0$ ($\lambda_w=0$) in Eq.(\ref{Louv1-eq}) and then evaluating the respective cumulant generating function and calculating their derivatives at $\lambda_q=0$ ($\lambda_w=0$). We numerically calculate the quantity  $\tilde\eta^{(2)}-(\tilde\eta^{(1)})^2$ for a wide range of engine parameters of the QHE. We find that Eq. (\ref{eq-saryal}) is not robust for the entire range of coherence parameters and varying efficiencies in our QHE as seen from Fig. (\ref{fig-bound_bj}).

We report that an analogue of the inequality in Eq.(\ref{eq-saryal}) exists between the first and second stochastic efficiency cumulants. Mathematically, 
\begin{align}
\label{eq-ineq}
    \eta^{(2)}-(\eta^{(1)})^2>0    
\end{align}
 We numerically evaluate the quantity  $\eta^{(2)}-(\eta^{(1)})^2$ for the same range of engine parameters as was done in Fig. (\ref{fig-bound_bj}). We find that Eq. (\ref{eq-ineq}) is robust for the entire range of coherence parameters and varying efficiencies as seen from Fig. (\ref{fig-bound_2}), where both positive and negative values are seen. However, no such bounds between the individual efficiency cumulants ($\eta^{(2)} \ and \ \eta^{(1)}$) exist as seen from the quantity $\eta^{(2)} - \eta^{(1)}$ in Fig. (\ref{fig-bound_2}) evaluated for a wide range of parameters. The negative values are observed when $|\eta^{(1)}| > \eta^{(2)}$. Negative values of $\eta^{(1)}$ are observed when the stochastic efficiency is close and equal to Carnot efficiency (Fig. \ref{fig2}(a)).


\section{Conclusion}
\label{conc}
In this work, we examined the first and second efficiency cumulants of a coherent four-level quantum heat engine. Firstly, we numerically computed the large deviation function for the stochastic efficiency of the QHE by deriving a generalized quantum master equation. We proved that the noise induced coherences resulting from the nonequlibrium nature of the engine do not influence the outcome of the two extremely likely (macroscopic) and unlikely (Carnot) efficiencies. We calculated the efficiency cumulants using a full counting statistics approach and showed that the value of the cumulants scaled by its classical value is independent of the stochastic efficiency for the entire range of hot and cold coherence parameters. We show this efficiency independence analytically near equilibrium for both the first and second efficiency cumulants. We also numerically demonstrated that, in the regime where the work is less than the thermal gradient, treating the efficiency as a stochastic variable allows the suppression of the second efficiency cumulant in the presence of coherences. However, this suppression does not occur in the absence of coherences. Additionally, we found that when the stochastic efficiency is at its most likely value (the engine's macroscopic quantum efficiency), there is no suppression of the second efficiency cumulant. We further discovered that the second efficiency cumulant is bounded below the square of the first efficiency cumulant irrespective of the engine parameters. This bound is analogous to a recently proposed bound between the second and the square of the first efficiency cumulants evaluated by separately calculating the work and heat fluctuations, which we found not to be robust in the considered cavity coupled quantum heat engine.

\begin{acknowledgments}
  MJS and HPG acknowledge the support from the Science and Engineering Research Board, India for the start-up grant, SERB/SRG/2021/001088. 
\end{acknowledgments}  




\bibliography{references.bib}

\begin{thebibliography}{49}%
\makeatletter
\providecommand \@ifxundefined [1]{%
 \@ifx{#1\undefined}
}%
\providecommand \@ifnum [1]{%
 \ifnum #1\expandafter \@firstoftwo
 \else \expandafter \@secondoftwo
 \fi
}%
\providecommand \@ifx [1]{%
 \ifx #1\expandafter \@firstoftwo
 \else \expandafter \@secondoftwo
 \fi
}%
\providecommand \natexlab [1]{#1}%
\providecommand \enquote  [1]{``#1''}%
\providecommand \bibnamefont  [1]{#1}%
\providecommand \bibfnamefont [1]{#1}%
\providecommand \citenamefont [1]{#1}%
\providecommand \href@noop [0]{\@secondoftwo}%
\providecommand \href [0]{\begingroup \@sanitize@url \@href}%
\providecommand \@href[1]{\@@startlink{#1}\@@href}%
\providecommand \@@href[1]{\endgroup#1\@@endlink}%
\providecommand \@sanitize@url [0]{\catcode `\\12\catcode `\$12\catcode `\&12\catcode `\#12\catcode `\^12\catcode `\_12\catcode `\%12\relax}%
\providecommand \@@startlink[1]{}%
\providecommand \@@endlink[0]{}%
\providecommand \url  [0]{\begingroup\@sanitize@url \@url }%
\providecommand \@url [1]{\endgroup\@href {#1}{\urlprefix }}%
\providecommand \urlprefix  [0]{URL }%
\providecommand \Eprint [0]{\href }%
\providecommand \doibase [0]{http://dx.doi.org/}%
\providecommand \selectlanguage [0]{\@gobble}%
\providecommand \bibinfo  [0]{\@secondoftwo}%
\providecommand \bibfield  [0]{\@secondoftwo}%
\providecommand \translation [1]{[#1]}%
\providecommand \BibitemOpen [0]{}%
\providecommand \bibitemStop [0]{}%
\providecommand \bibitemNoStop [0]{.\EOS\space}%
\providecommand \EOS [0]{\spacefactor3000\relax}%
\providecommand \BibitemShut  [1]{\csname bibitem#1\endcsname}%
\let\auto@bib@innerbib\@empty
\bibitem [{\citenamefont {Gingrich}\ \emph {et~al.}(2014)\citenamefont {Gingrich}, \citenamefont {Rotskoff}, \citenamefont {Vaikuntanathan},\ and\ \citenamefont {Geissler}}]{Gingrich_2014Efficiency_large_deviations}%
  \BibitemOpen
  \bibfield  {author} {\bibinfo {author} {\bibfnamefont {T.~R.}\ \bibnamefont {Gingrich}}, \bibinfo {author} {\bibfnamefont {G.~M.}\ \bibnamefont {Rotskoff}}, \bibinfo {author} {\bibfnamefont {S.}~\bibnamefont {Vaikuntanathan}}, \ and\ \bibinfo {author} {\bibfnamefont {P.~L.}\ \bibnamefont {Geissler}},\ }\href {\doibase 10.1088/1367-2630/16/10/102003} {\bibfield  {journal} {\bibinfo  {journal} {New Journal of Physics}\ }\textbf {\bibinfo {volume} {16}},\ \bibinfo {pages} {102003} (\bibinfo {year} {2014})}\BibitemShut {NoStop}%
\bibitem [{\citenamefont {Verley}\ \emph {et~al.}(2014{\natexlab{a}})\citenamefont {Verley}, \citenamefont {Willaert}, \citenamefont {Van~den Broeck},\ and\ \citenamefont {Esposito}}]{PhysRevE.90.052145_Universal_theory_of_efficiency}%
  \BibitemOpen
  \bibfield  {author} {\bibinfo {author} {\bibfnamefont {G.}~\bibnamefont {Verley}}, \bibinfo {author} {\bibfnamefont {T.}~\bibnamefont {Willaert}}, \bibinfo {author} {\bibfnamefont {C.}~\bibnamefont {Van~den Broeck}}, \ and\ \bibinfo {author} {\bibfnamefont {M.}~\bibnamefont {Esposito}},\ }\href {\doibase 10.1103/PhysRevE.90.052145} {\bibfield  {journal} {\bibinfo  {journal} {Phys. Rev. E}\ }\textbf {\bibinfo {volume} {90}},\ \bibinfo {pages} {052145} (\bibinfo {year} {2014}{\natexlab{a}})}\BibitemShut {NoStop}%
\bibitem [{\citenamefont {Vroylandt}\ \emph {et~al.}(2016)\citenamefont {Vroylandt}, \citenamefont {Bonfils},\ and\ \citenamefont {Verley}}]{PhysRevE.93.052123_eff_flu_small_machines}%
  \BibitemOpen
  \bibfield  {author} {\bibinfo {author} {\bibfnamefont {H.}~\bibnamefont {Vroylandt}}, \bibinfo {author} {\bibfnamefont {A.}~\bibnamefont {Bonfils}}, \ and\ \bibinfo {author} {\bibfnamefont {G.}~\bibnamefont {Verley}},\ }\href {\doibase 10.1103/PhysRevE.93.052123} {\bibfield  {journal} {\bibinfo  {journal} {Phys. Rev. E}\ }\textbf {\bibinfo {volume} {93}},\ \bibinfo {pages} {052123} (\bibinfo {year} {2016})}\BibitemShut {NoStop}%
\bibitem [{\citenamefont {Polettini}\ \emph {et~al.}(2015)\citenamefont {Polettini}, \citenamefont {Verley},\ and\ \citenamefont {Esposito}}]{PhysRevLett.114.050601Efficiency_Statistics_at_All}%
  \BibitemOpen
  \bibfield  {author} {\bibinfo {author} {\bibfnamefont {M.}~\bibnamefont {Polettini}}, \bibinfo {author} {\bibfnamefont {G.}~\bibnamefont {Verley}}, \ and\ \bibinfo {author} {\bibfnamefont {M.}~\bibnamefont {Esposito}},\ }\href {\doibase 10.1103/PhysRevLett.114.050601} {\bibfield  {journal} {\bibinfo  {journal} {Phys. Rev. Lett.}\ }\textbf {\bibinfo {volume} {114}},\ \bibinfo {pages} {050601} (\bibinfo {year} {2015})}\BibitemShut {NoStop}%
\bibitem [{\citenamefont {Manikandan}\ \emph {et~al.}(2019)\citenamefont {Manikandan}, \citenamefont {Dabelow}, \citenamefont {Eichhorn},\ and\ \citenamefont {Krishnamurthy}}]{PhysRevLett.122.140601_eff_flu_micro_mac}%
  \BibitemOpen
  \bibfield  {author} {\bibinfo {author} {\bibfnamefont {S.~K.}\ \bibnamefont {Manikandan}}, \bibinfo {author} {\bibfnamefont {L.}~\bibnamefont {Dabelow}}, \bibinfo {author} {\bibfnamefont {R.}~\bibnamefont {Eichhorn}}, \ and\ \bibinfo {author} {\bibfnamefont {S.}~\bibnamefont {Krishnamurthy}},\ }\href {\doibase 10.1103/PhysRevLett.122.140601} {\bibfield  {journal} {\bibinfo  {journal} {Phys. Rev. Lett.}\ }\textbf {\bibinfo {volume} {122}},\ \bibinfo {pages} {140601} (\bibinfo {year} {2019})}\BibitemShut {NoStop}%
\bibitem [{\citenamefont {Verley}\ \emph {et~al.}(2014{\natexlab{b}})\citenamefont {Verley}, \citenamefont {Esposito}, \citenamefont {Willaert},\ and\ \citenamefont {Van~den Broeck}}]{Verley2014_unlikely_Carnot_efficiency}%
  \BibitemOpen
  \bibfield  {author} {\bibinfo {author} {\bibfnamefont {G.}~\bibnamefont {Verley}}, \bibinfo {author} {\bibfnamefont {M.}~\bibnamefont {Esposito}}, \bibinfo {author} {\bibfnamefont {T.}~\bibnamefont {Willaert}}, \ and\ \bibinfo {author} {\bibfnamefont {C.}~\bibnamefont {Van~den Broeck}},\ }\href {\doibase 10.1038/ncomms5721} {\bibfield  {journal} {\bibinfo  {journal} {Nature Communications}\ }\textbf {\bibinfo {volume} {5}},\ \bibinfo {pages} {4721} (\bibinfo {year} {2014}{\natexlab{b}})}\BibitemShut {NoStop}%
\bibitem [{\citenamefont {Mart{\'\i}nez}\ \emph {et~al.}(2016)\citenamefont {Mart{\'\i}nez}, \citenamefont {Rold{\'a}n}, \citenamefont {Dinis}, \citenamefont {Petrov}, \citenamefont {Parrondo},\ and\ \citenamefont {Rica}}]{martinez2016_brownian_carnot_engine}%
  \BibitemOpen
  \bibfield  {author} {\bibinfo {author} {\bibfnamefont {I.~A.}\ \bibnamefont {Mart{\'\i}nez}}, \bibinfo {author} {\bibfnamefont {{\'E}.}~\bibnamefont {Rold{\'a}n}}, \bibinfo {author} {\bibfnamefont {L.}~\bibnamefont {Dinis}}, \bibinfo {author} {\bibfnamefont {D.}~\bibnamefont {Petrov}}, \bibinfo {author} {\bibfnamefont {J.~M.}\ \bibnamefont {Parrondo}}, \ and\ \bibinfo {author} {\bibfnamefont {R.~A.}\ \bibnamefont {Rica}},\ }\href@noop {} {\bibfield  {journal} {\bibinfo  {journal} {Nature physics}\ }\textbf {\bibinfo {volume} {12}},\ \bibinfo {pages} {67} (\bibinfo {year} {2016})}\BibitemShut {NoStop}%
\bibitem [{\citenamefont {Deffner}\ and\ \citenamefont {Lutz}(2008)}]{PhysRevE.77.021128_non_eq_work_dis}%
  \BibitemOpen
  \bibfield  {author} {\bibinfo {author} {\bibfnamefont {S.}~\bibnamefont {Deffner}}\ and\ \bibinfo {author} {\bibfnamefont {E.}~\bibnamefont {Lutz}},\ }\href {\doibase 10.1103/PhysRevE.77.021128} {\bibfield  {journal} {\bibinfo  {journal} {Phys. Rev. E}\ }\textbf {\bibinfo {volume} {77}},\ \bibinfo {pages} {021128} (\bibinfo {year} {2008})}\BibitemShut {NoStop}%
\bibitem [{\citenamefont {Huber}\ \emph {et~al.}(2008)\citenamefont {Huber}, \citenamefont {Schmidt-Kaler}, \citenamefont {Deffner},\ and\ \citenamefont {Lutz}}]{PhysRevLett.101.070403_Trapped_Cold_atom_emp}%
  \BibitemOpen
  \bibfield  {author} {\bibinfo {author} {\bibfnamefont {G.}~\bibnamefont {Huber}}, \bibinfo {author} {\bibfnamefont {F.}~\bibnamefont {Schmidt-Kaler}}, \bibinfo {author} {\bibfnamefont {S.}~\bibnamefont {Deffner}}, \ and\ \bibinfo {author} {\bibfnamefont {E.}~\bibnamefont {Lutz}},\ }\href {\doibase 10.1103/PhysRevLett.101.070403} {\bibfield  {journal} {\bibinfo  {journal} {Phys. Rev. Lett.}\ }\textbf {\bibinfo {volume} {101}},\ \bibinfo {pages} {070403} (\bibinfo {year} {2008})}\BibitemShut {NoStop}%
\bibitem [{\citenamefont {Talkner}\ \emph {et~al.}(2008)\citenamefont {Talkner}, \citenamefont {Burada},\ and\ \citenamefont {H\"anggi}}]{PhysRevE.78.011115_forced_qun_osc}%
  \BibitemOpen
  \bibfield  {author} {\bibinfo {author} {\bibfnamefont {P.}~\bibnamefont {Talkner}}, \bibinfo {author} {\bibfnamefont {P.~S.}\ \bibnamefont {Burada}}, \ and\ \bibinfo {author} {\bibfnamefont {P.}~\bibnamefont {H\"anggi}},\ }\href {\doibase 10.1103/PhysRevE.78.011115} {\bibfield  {journal} {\bibinfo  {journal} {Phys. Rev. E}\ }\textbf {\bibinfo {volume} {78}},\ \bibinfo {pages} {011115} (\bibinfo {year} {2008})}\BibitemShut {NoStop}%
\bibitem [{\citenamefont {Solinas}\ \emph {et~al.}(2013)\citenamefont {Solinas}, \citenamefont {Averin},\ and\ \citenamefont {Pekola}}]{PhysRevB.87.060508_fluc_dri_q_sys}%
  \BibitemOpen
  \bibfield  {author} {\bibinfo {author} {\bibfnamefont {P.}~\bibnamefont {Solinas}}, \bibinfo {author} {\bibfnamefont {D.~V.}\ \bibnamefont {Averin}}, \ and\ \bibinfo {author} {\bibfnamefont {J.~P.}\ \bibnamefont {Pekola}},\ }\href {\doibase 10.1103/PhysRevB.87.060508} {\bibfield  {journal} {\bibinfo  {journal} {Phys. Rev. B}\ }\textbf {\bibinfo {volume} {87}},\ \bibinfo {pages} {060508} (\bibinfo {year} {2013})}\BibitemShut {NoStop}%
\bibitem [{\citenamefont {Hekking}\ and\ \citenamefont {Pekola}(2013)}]{PhysRevLett.111.093602_w_didd_2_l_sys}%
  \BibitemOpen
  \bibfield  {author} {\bibinfo {author} {\bibfnamefont {F.~W.~J.}\ \bibnamefont {Hekking}}\ and\ \bibinfo {author} {\bibfnamefont {J.~P.}\ \bibnamefont {Pekola}},\ }\href {\doibase 10.1103/PhysRevLett.111.093602} {\bibfield  {journal} {\bibinfo  {journal} {Phys. Rev. Lett.}\ }\textbf {\bibinfo {volume} {111}},\ \bibinfo {pages} {093602} (\bibinfo {year} {2013})}\BibitemShut {NoStop}%
\bibitem [{\citenamefont {An}\ \emph {et~al.}(2015)\citenamefont {An}, \citenamefont {Zhang}, \citenamefont {Um}, \citenamefont {Lv}, \citenamefont {Lu}, \citenamefont {Zhang}, \citenamefont {Yin}, \citenamefont {Quan},\ and\ \citenamefont {Kim}}]{An2015_ion_trap}%
  \BibitemOpen
  \bibfield  {author} {\bibinfo {author} {\bibfnamefont {S.}~\bibnamefont {An}}, \bibinfo {author} {\bibfnamefont {J.-N.}\ \bibnamefont {Zhang}}, \bibinfo {author} {\bibfnamefont {M.}~\bibnamefont {Um}}, \bibinfo {author} {\bibfnamefont {D.}~\bibnamefont {Lv}}, \bibinfo {author} {\bibfnamefont {Y.}~\bibnamefont {Lu}}, \bibinfo {author} {\bibfnamefont {J.}~\bibnamefont {Zhang}}, \bibinfo {author} {\bibfnamefont {Z.-Q.}\ \bibnamefont {Yin}}, \bibinfo {author} {\bibfnamefont {H.~T.}\ \bibnamefont {Quan}}, \ and\ \bibinfo {author} {\bibfnamefont {K.}~\bibnamefont {Kim}},\ }\href {\doibase 10.1038/nphys3197} {\bibfield  {journal} {\bibinfo  {journal} {Nature Physics}\ }\textbf {\bibinfo {volume} {11}},\ \bibinfo {pages} {193} (\bibinfo {year} {2015})}\BibitemShut {NoStop}%
\bibitem [{\citenamefont {Batalh\~ao}\ \emph {et~al.}(2014)\citenamefont {Batalh\~ao}, \citenamefont {Souza}, \citenamefont {Mazzola}, \citenamefont {Auccaise}, \citenamefont {Sarthour}, \citenamefont {Oliveira}, \citenamefont {Goold}, \citenamefont {De~Chiara}, \citenamefont {Paternostro},\ and\ \citenamefont {Serra}}]{PhysRevLett.113.140601ExperimentalReconstructionWorkDistribution}%
  \BibitemOpen
  \bibfield  {author} {\bibinfo {author} {\bibfnamefont {T.~B.}\ \bibnamefont {Batalh\~ao}}, \bibinfo {author} {\bibfnamefont {A.~M.}\ \bibnamefont {Souza}}, \bibinfo {author} {\bibfnamefont {L.}~\bibnamefont {Mazzola}}, \bibinfo {author} {\bibfnamefont {R.}~\bibnamefont {Auccaise}}, \bibinfo {author} {\bibfnamefont {R.~S.}\ \bibnamefont {Sarthour}}, \bibinfo {author} {\bibfnamefont {I.~S.}\ \bibnamefont {Oliveira}}, \bibinfo {author} {\bibfnamefont {J.}~\bibnamefont {Goold}}, \bibinfo {author} {\bibfnamefont {G.}~\bibnamefont {De~Chiara}}, \bibinfo {author} {\bibfnamefont {M.}~\bibnamefont {Paternostro}}, \ and\ \bibinfo {author} {\bibfnamefont {R.~M.}\ \bibnamefont {Serra}},\ }\href {\doibase 10.1103/PhysRevLett.113.140601} {\bibfield  {journal} {\bibinfo  {journal} {Phys. Rev. Lett.}\ }\textbf {\bibinfo {volume} {113}},\ \bibinfo {pages} {140601} (\bibinfo {year} {2014})}\BibitemShut {NoStop}%
\bibitem [{\citenamefont {Cerisola}\ \emph {et~al.}(2017)\citenamefont {Cerisola}, \citenamefont {Margalit}, \citenamefont {Machluf}, \citenamefont {Roncaglia}, \citenamefont {Paz},\ and\ \citenamefont {Folman}}]{Cerisola2017_colsAtom_setups}%
  \BibitemOpen
  \bibfield  {author} {\bibinfo {author} {\bibfnamefont {F.}~\bibnamefont {Cerisola}}, \bibinfo {author} {\bibfnamefont {Y.}~\bibnamefont {Margalit}}, \bibinfo {author} {\bibfnamefont {S.}~\bibnamefont {Machluf}}, \bibinfo {author} {\bibfnamefont {A.~J.}\ \bibnamefont {Roncaglia}}, \bibinfo {author} {\bibfnamefont {J.~P.}\ \bibnamefont {Paz}}, \ and\ \bibinfo {author} {\bibfnamefont {R.}~\bibnamefont {Folman}},\ }\href {\doibase 10.1038/s41467-017-01308-7} {\bibfield  {journal} {\bibinfo  {journal} {Nature Communications}\ }\textbf {\bibinfo {volume} {8}},\ \bibinfo {pages} {1241} (\bibinfo {year} {2017})}\BibitemShut {NoStop}%
\bibitem [{\citenamefont {Esposito}\ \emph {et~al.}(2015)\citenamefont {Esposito}, \citenamefont {Ochoa},\ and\ \citenamefont {Galperin}}]{PhysRevB.91.115417_eff_fluc_q_thermo_ele_devices}%
  \BibitemOpen
  \bibfield  {author} {\bibinfo {author} {\bibfnamefont {M.}~\bibnamefont {Esposito}}, \bibinfo {author} {\bibfnamefont {M.~A.}\ \bibnamefont {Ochoa}}, \ and\ \bibinfo {author} {\bibfnamefont {M.}~\bibnamefont {Galperin}},\ }\href {\doibase 10.1103/PhysRevB.91.115417} {\bibfield  {journal} {\bibinfo  {journal} {Phys. Rev. B}\ }\textbf {\bibinfo {volume} {91}},\ \bibinfo {pages} {115417} (\bibinfo {year} {2015})}\BibitemShut {NoStop}%
\bibitem [{\citenamefont {Campisi}\ \emph {et~al.}(2015)\citenamefont {Campisi}, \citenamefont {Pekola},\ and\ \citenamefont {Fazio}}]{Campisi_2015_egs_swap}%
  \BibitemOpen
  \bibfield  {author} {\bibinfo {author} {\bibfnamefont {M.}~\bibnamefont {Campisi}}, \bibinfo {author} {\bibfnamefont {J.}~\bibnamefont {Pekola}}, \ and\ \bibinfo {author} {\bibfnamefont {R.}~\bibnamefont {Fazio}},\ }\href {\doibase 10.1088/1367-2630/17/3/035012} {\bibfield  {journal} {\bibinfo  {journal} {New Journal of Physics}\ }\textbf {\bibinfo {volume} {17}},\ \bibinfo {pages} {035012} (\bibinfo {year} {2015})}\BibitemShut {NoStop}%
\bibitem [{\citenamefont {Callen}(1991)}]{book_thermodynmics}%
  \BibitemOpen
  \bibfield  {author} {\bibinfo {author} {\bibfnamefont {H.~B.}\ \bibnamefont {Callen}},\ }\href@noop {} {\enquote {\bibinfo {title} {Thermodynamics and an introduction to thermostatistics, 2nd edition},}\ } (\bibinfo {year} {Jan 1, 1991})\BibitemShut {NoStop}%
\bibitem [{\citenamefont {Sinitsyn}(2011)}]{Sinitsyn_2011_FT_1}%
  \BibitemOpen
  \bibfield  {author} {\bibinfo {author} {\bibfnamefont {N.~A.}\ \bibnamefont {Sinitsyn}},\ }\href {\doibase 10.1088/1751-8113/44/40/405001} {\bibfield  {journal} {\bibinfo  {journal} {Journal of Physics A: Mathematical and Theoretical}\ }\textbf {\bibinfo {volume} {44}},\ \bibinfo {pages} {405001} (\bibinfo {year} {2011})}\BibitemShut {NoStop}%
\bibitem [{\citenamefont {Campisi}(2014)}]{Campisi_2014_FT2}%
  \BibitemOpen
  \bibfield  {author} {\bibinfo {author} {\bibfnamefont {M.}~\bibnamefont {Campisi}},\ }\href {\doibase 10.1088/1751-8113/47/24/245001} {\bibfield  {journal} {\bibinfo  {journal} {Journal of Physics A: Mathematical and Theoretical}\ }\textbf {\bibinfo {volume} {47}},\ \bibinfo {pages} {245001} (\bibinfo {year} {2014})}\BibitemShut {NoStop}%
\bibitem [{\citenamefont {Rao}\ and\ \citenamefont {Esposito}(2018)}]{e20090635_FT3}%
  \BibitemOpen
  \bibfield  {author} {\bibinfo {author} {\bibfnamefont {R.}~\bibnamefont {Rao}}\ and\ \bibinfo {author} {\bibfnamefont {M.}~\bibnamefont {Esposito}},\ }\href {\doibase 10.3390/e20090635} {\bibfield  {journal} {\bibinfo  {journal} {Entropy}\ }\textbf {\bibinfo {volume} {20}} (\bibinfo {year} {2018}),\ 10.3390/e20090635}\BibitemShut {NoStop}%
\bibitem [{\citenamefont {Proesmans}\ \emph {et~al.}(2015{\natexlab{a}})\citenamefont {Proesmans}, \citenamefont {Driesen}, \citenamefont {Cleuren},\ and\ \citenamefont {Van~den Broeck}}]{PhysRevE.92.032105_eff_single_particle_engines}%
  \BibitemOpen
  \bibfield  {author} {\bibinfo {author} {\bibfnamefont {K.}~\bibnamefont {Proesmans}}, \bibinfo {author} {\bibfnamefont {C.}~\bibnamefont {Driesen}}, \bibinfo {author} {\bibfnamefont {B.}~\bibnamefont {Cleuren}}, \ and\ \bibinfo {author} {\bibfnamefont {C.}~\bibnamefont {Van~den Broeck}},\ }\href {\doibase 10.1103/PhysRevE.92.032105} {\bibfield  {journal} {\bibinfo  {journal} {Phys. Rev. E}\ }\textbf {\bibinfo {volume} {92}},\ \bibinfo {pages} {032105} (\bibinfo {year} {2015}{\natexlab{a}})}\BibitemShut {NoStop}%
\bibitem [{\citenamefont {Proesmans}\ \emph {et~al.}(2015{\natexlab{b}})\citenamefont {Proesmans}, \citenamefont {Cleuren},\ and\ \citenamefont {den Broeck}}]{Proesmans_2015_effi_fluc_effusion}%
  \BibitemOpen
  \bibfield  {author} {\bibinfo {author} {\bibfnamefont {K.}~\bibnamefont {Proesmans}}, \bibinfo {author} {\bibfnamefont {B.}~\bibnamefont {Cleuren}}, \ and\ \bibinfo {author} {\bibfnamefont {C.~V.}\ \bibnamefont {den Broeck}},\ }\href {\doibase 10.1209/0295-5075/109/20004} {\bibfield  {journal} {\bibinfo  {journal} {Europhysics Letters}\ }\textbf {\bibinfo {volume} {109}},\ \bibinfo {pages} {20004} (\bibinfo {year} {2015}{\natexlab{b}})}\BibitemShut {NoStop}%
\bibitem [{\citenamefont {Jiang}\ \emph {et~al.}(2015)\citenamefont {Jiang}, \citenamefont {Agarwalla},\ and\ \citenamefont {Segal}}]{PhysRevLett.115.040601_eff_fluct_broken_time_rev_sym}%
  \BibitemOpen
  \bibfield  {author} {\bibinfo {author} {\bibfnamefont {J.-H.}\ \bibnamefont {Jiang}}, \bibinfo {author} {\bibfnamefont {B.~K.}\ \bibnamefont {Agarwalla}}, \ and\ \bibinfo {author} {\bibfnamefont {D.}~\bibnamefont {Segal}},\ }\href {\doibase 10.1103/PhysRevLett.115.040601} {\bibfield  {journal} {\bibinfo  {journal} {Phys. Rev. Lett.}\ }\textbf {\bibinfo {volume} {115}},\ \bibinfo {pages} {040601} (\bibinfo {year} {2015})}\BibitemShut {NoStop}%
\bibitem [{\citenamefont {Agarwalla}\ \emph {et~al.}(2015{\natexlab{a}})\citenamefont {Agarwalla}, \citenamefont {Jiang},\ and\ \citenamefont {Segal}}]{PhysRevB.92.245418_fcs_vibration}%
  \BibitemOpen
  \bibfield  {author} {\bibinfo {author} {\bibfnamefont {B.~K.}\ \bibnamefont {Agarwalla}}, \bibinfo {author} {\bibfnamefont {J.-H.}\ \bibnamefont {Jiang}}, \ and\ \bibinfo {author} {\bibfnamefont {D.}~\bibnamefont {Segal}},\ }\href {\doibase 10.1103/PhysRevB.92.245418} {\bibfield  {journal} {\bibinfo  {journal} {Phys. Rev. B}\ }\textbf {\bibinfo {volume} {92}},\ \bibinfo {pages} {245418} (\bibinfo {year} {2015}{\natexlab{a}})}\BibitemShut {NoStop}%
\bibitem [{\citenamefont {Proesmans}\ \emph {et~al.}(2016)\citenamefont {Proesmans}, \citenamefont {Dreher}, \citenamefont {Gavrilov}, \citenamefont {Bechhoefer},\ and\ \citenamefont {Van~den Broeck}}]{PhysRevX.6.041010_brownian_duet}%
  \BibitemOpen
  \bibfield  {author} {\bibinfo {author} {\bibfnamefont {K.}~\bibnamefont {Proesmans}}, \bibinfo {author} {\bibfnamefont {Y.}~\bibnamefont {Dreher}}, \bibinfo {author} {\bibfnamefont {M.~c.~v.}\ \bibnamefont {Gavrilov}}, \bibinfo {author} {\bibfnamefont {J.}~\bibnamefont {Bechhoefer}}, \ and\ \bibinfo {author} {\bibfnamefont {C.}~\bibnamefont {Van~den Broeck}},\ }\href {\doibase 10.1103/PhysRevX.6.041010} {\bibfield  {journal} {\bibinfo  {journal} {Phys. Rev. X}\ }\textbf {\bibinfo {volume} {6}},\ \bibinfo {pages} {041010} (\bibinfo {year} {2016})}\BibitemShut {NoStop}%
\bibitem [{\citenamefont {Allahverdyan}\ and\ \citenamefont {Hovhannisyan}(2011)}]{allahverdyan2011work_single}%
  \BibitemOpen
  \bibfield  {author} {\bibinfo {author} {\bibfnamefont {A.~E.}\ \bibnamefont {Allahverdyan}}\ and\ \bibinfo {author} {\bibfnamefont {K.~V.}\ \bibnamefont {Hovhannisyan}},\ }\href@noop {} {\bibfield  {journal} {\bibinfo  {journal} {Europhysics Letters}\ }\textbf {\bibinfo {volume} {95}},\ \bibinfo {pages} {60004} (\bibinfo {year} {2011})}\BibitemShut {NoStop}%
\bibitem [{\citenamefont {Scully}\ \emph {et~al.}(2003)\citenamefont {Scully}, \citenamefont {Zubairy}, \citenamefont {Agarwal},\ and\ \citenamefont {Walther}}]{scully2003extracting_single}%
  \BibitemOpen
  \bibfield  {author} {\bibinfo {author} {\bibfnamefont {M.~O.}\ \bibnamefont {Scully}}, \bibinfo {author} {\bibfnamefont {M.~S.}\ \bibnamefont {Zubairy}}, \bibinfo {author} {\bibfnamefont {G.~S.}\ \bibnamefont {Agarwal}}, \ and\ \bibinfo {author} {\bibfnamefont {H.}~\bibnamefont {Walther}},\ }\href@noop {} {\bibfield  {journal} {\bibinfo  {journal} {Science}\ }\textbf {\bibinfo {volume} {299}},\ \bibinfo {pages} {862} (\bibinfo {year} {2003})}\BibitemShut {NoStop}%
\bibitem [{\citenamefont {Scully}(2010)}]{PhysRevLett.104.207701_photocell}%
  \BibitemOpen
  \bibfield  {author} {\bibinfo {author} {\bibfnamefont {M.~O.}\ \bibnamefont {Scully}},\ }\href {\doibase 10.1103/PhysRevLett.104.207701} {\bibfield  {journal} {\bibinfo  {journal} {Phys. Rev. Lett.}\ }\textbf {\bibinfo {volume} {104}},\ \bibinfo {pages} {207701} (\bibinfo {year} {2010})}\BibitemShut {NoStop}%
\bibitem [{\citenamefont {Kilin}\ \emph {et~al.}(2008)\citenamefont {Kilin}, \citenamefont {Kapale},\ and\ \citenamefont {Scully}}]{PhysRevLett.100.173601_lasing}%
  \BibitemOpen
  \bibfield  {author} {\bibinfo {author} {\bibfnamefont {S.~Y.}\ \bibnamefont {Kilin}}, \bibinfo {author} {\bibfnamefont {K.~T.}\ \bibnamefont {Kapale}}, \ and\ \bibinfo {author} {\bibfnamefont {M.~O.}\ \bibnamefont {Scully}},\ }\href {\doibase 10.1103/PhysRevLett.100.173601} {\bibfield  {journal} {\bibinfo  {journal} {Phys. Rev. Lett.}\ }\textbf {\bibinfo {volume} {100}},\ \bibinfo {pages} {173601} (\bibinfo {year} {2008})}\BibitemShut {NoStop}%
\bibitem [{\citenamefont {Rahav}\ \emph {et~al.}(2012{\natexlab{a}})\citenamefont {Rahav}, \citenamefont {Harbola},\ and\ \citenamefont {Mukamel}}]{PhysRevA.86.043843_rahav}%
  \BibitemOpen
  \bibfield  {author} {\bibinfo {author} {\bibfnamefont {S.}~\bibnamefont {Rahav}}, \bibinfo {author} {\bibfnamefont {U.}~\bibnamefont {Harbola}}, \ and\ \bibinfo {author} {\bibfnamefont {S.}~\bibnamefont {Mukamel}},\ }\href {\doibase 10.1103/PhysRevA.86.043843} {\bibfield  {journal} {\bibinfo  {journal} {Phys. Rev. A}\ }\textbf {\bibinfo {volume} {86}},\ \bibinfo {pages} {043843} (\bibinfo {year} {2012}{\natexlab{a}})}\BibitemShut {NoStop}%
\bibitem [{\citenamefont {Scully}\ \emph {et~al.}(2011)\citenamefont {Scully}, \citenamefont {Chapin}, \citenamefont {Dorfman}, \citenamefont {Kim},\ and\ \citenamefont {Svidzinsky}}]{scully2011quantum_noise}%
  \BibitemOpen
  \bibfield  {author} {\bibinfo {author} {\bibfnamefont {M.~O.}\ \bibnamefont {Scully}}, \bibinfo {author} {\bibfnamefont {K.~R.}\ \bibnamefont {Chapin}}, \bibinfo {author} {\bibfnamefont {K.~E.}\ \bibnamefont {Dorfman}}, \bibinfo {author} {\bibfnamefont {M.~B.}\ \bibnamefont {Kim}}, \ and\ \bibinfo {author} {\bibfnamefont {A.}~\bibnamefont {Svidzinsky}},\ }\href@noop {} {\bibfield  {journal} {\bibinfo  {journal} {Proceedings of the National Academy of Sciences}\ }\textbf {\bibinfo {volume} {108}},\ \bibinfo {pages} {15097} (\bibinfo {year} {2011})}\BibitemShut {NoStop}%
\bibitem [{\citenamefont {Camati}\ \emph {et~al.}(2019)\citenamefont {Camati}, \citenamefont {Santos},\ and\ \citenamefont {Serra}}]{PhysRevA.99.062103_otto_coh}%
  \BibitemOpen
  \bibfield  {author} {\bibinfo {author} {\bibfnamefont {P.~A.}\ \bibnamefont {Camati}}, \bibinfo {author} {\bibfnamefont {J.~F.~G.}\ \bibnamefont {Santos}}, \ and\ \bibinfo {author} {\bibfnamefont {R.~M.}\ \bibnamefont {Serra}},\ }\href {\doibase 10.1103/PhysRevA.99.062103} {\bibfield  {journal} {\bibinfo  {journal} {Phys. Rev. A}\ }\textbf {\bibinfo {volume} {99}},\ \bibinfo {pages} {062103} (\bibinfo {year} {2019})}\BibitemShut {NoStop}%
\bibitem [{\citenamefont {Sarmah}\ and\ \citenamefont {Goswami}(2023{\natexlab{a}})}]{sarmah2023learning}%
  \BibitemOpen
  \bibfield  {author} {\bibinfo {author} {\bibfnamefont {M.~J.}\ \bibnamefont {Sarmah}}\ and\ \bibinfo {author} {\bibfnamefont {H.~P.}\ \bibnamefont {Goswami}},\ }\href@noop {} {\bibfield  {journal} {\bibinfo  {journal} {Physica A: Statistical Mechanics and its Applications}\ }\textbf {\bibinfo {volume} {627}},\ \bibinfo {pages} {129135} (\bibinfo {year} {2023}{\natexlab{a}})}\BibitemShut {NoStop}%
\bibitem [{\citenamefont {Sarmah}\ and\ \citenamefont {Goswami}(2023{\natexlab{b}})}]{PhysRevA.107.052217_mjs}%
  \BibitemOpen
  \bibfield  {author} {\bibinfo {author} {\bibfnamefont {M.~J.}\ \bibnamefont {Sarmah}}\ and\ \bibinfo {author} {\bibfnamefont {H.~P.}\ \bibnamefont {Goswami}},\ }\href {\doibase 10.1103/PhysRevA.107.052217} {\bibfield  {journal} {\bibinfo  {journal} {Phys. Rev. A}\ }\textbf {\bibinfo {volume} {107}},\ \bibinfo {pages} {052217} (\bibinfo {year} {2023}{\natexlab{b}})}\BibitemShut {NoStop}%
\bibitem [{\citenamefont {Huang}\ \emph {et~al.}(2017)\citenamefont {Huang}, \citenamefont {Zou}, \citenamefont {Mao}, \citenamefont {Corp}, \citenamefont {Yao}, \citenamefont {Lee}, \citenamefont {Schlenker}, \citenamefont {Jen},\ and\ \citenamefont {Lin}}]{zou2017quantum}%
  \BibitemOpen
  \bibfield  {author} {\bibinfo {author} {\bibfnamefont {C.-Y.}\ \bibnamefont {Huang}}, \bibinfo {author} {\bibfnamefont {C.}~\bibnamefont {Zou}}, \bibinfo {author} {\bibfnamefont {C.}~\bibnamefont {Mao}}, \bibinfo {author} {\bibfnamefont {K.~L.}\ \bibnamefont {Corp}}, \bibinfo {author} {\bibfnamefont {Y.-C.}\ \bibnamefont {Yao}}, \bibinfo {author} {\bibfnamefont {Y.-J.}\ \bibnamefont {Lee}}, \bibinfo {author} {\bibfnamefont {C.~W.}\ \bibnamefont {Schlenker}}, \bibinfo {author} {\bibfnamefont {A.~K.~Y.}\ \bibnamefont {Jen}}, \ and\ \bibinfo {author} {\bibfnamefont {L.~Y.}\ \bibnamefont {Lin}},\ }\href {\doibase 10.1021/acsphotonics.7b00520} {\bibfield  {journal} {\bibinfo  {journal} {ACS Photonics}\ }\textbf {\bibinfo {volume} {4}},\ \bibinfo {pages} {2281} (\bibinfo {year} {2017})}\BibitemShut {NoStop}%
\bibitem [{\citenamefont {Bouton}\ \emph {et~al.}(2021)\citenamefont {Bouton}, \citenamefont {Nettersheim}, \citenamefont {Burgardt}, \citenamefont {Adam}, \citenamefont {Lutz},\ and\ \citenamefont {Widera}}]{bouton2021quantum}%
  \BibitemOpen
  \bibfield  {author} {\bibinfo {author} {\bibfnamefont {Q.}~\bibnamefont {Bouton}}, \bibinfo {author} {\bibfnamefont {J.}~\bibnamefont {Nettersheim}}, \bibinfo {author} {\bibfnamefont {S.}~\bibnamefont {Burgardt}}, \bibinfo {author} {\bibfnamefont {D.}~\bibnamefont {Adam}}, \bibinfo {author} {\bibfnamefont {E.}~\bibnamefont {Lutz}}, \ and\ \bibinfo {author} {\bibfnamefont {A.}~\bibnamefont {Widera}},\ }\href@noop {} {\bibfield  {journal} {\bibinfo  {journal} {Nature Communications}\ }\textbf {\bibinfo {volume} {12}},\ \bibinfo {pages} {2063} (\bibinfo {year} {2021})}\BibitemShut {NoStop}%
\bibitem [{\citenamefont {Esposito}\ \emph {et~al.}(2009{\natexlab{a}})\citenamefont {Esposito}, \citenamefont {Harbola},\ and\ \citenamefont {Mukamel}}]{RevModPhys.81.1665_harbola_rev}%
  \BibitemOpen
  \bibfield  {author} {\bibinfo {author} {\bibfnamefont {M.}~\bibnamefont {Esposito}}, \bibinfo {author} {\bibfnamefont {U.}~\bibnamefont {Harbola}}, \ and\ \bibinfo {author} {\bibfnamefont {S.}~\bibnamefont {Mukamel}},\ }\href {\doibase 10.1103/RevModPhys.81.1665} {\bibfield  {journal} {\bibinfo  {journal} {Rev. Mod. Phys.}\ }\textbf {\bibinfo {volume} {81}},\ \bibinfo {pages} {1665} (\bibinfo {year} {2009}{\natexlab{a}})}\BibitemShut {NoStop}%
\bibitem [{\citenamefont {Agarwalla}\ \emph {et~al.}(2015{\natexlab{b}})\citenamefont {Agarwalla}, \citenamefont {Jiang},\ and\ \citenamefont {Segal}}]{PhysRevB.92.245418_bijay_prb2015}%
  \BibitemOpen
  \bibfield  {author} {\bibinfo {author} {\bibfnamefont {B.~K.}\ \bibnamefont {Agarwalla}}, \bibinfo {author} {\bibfnamefont {J.-H.}\ \bibnamefont {Jiang}}, \ and\ \bibinfo {author} {\bibfnamefont {D.}~\bibnamefont {Segal}},\ }\href {\doibase 10.1103/PhysRevB.92.245418} {\bibfield  {journal} {\bibinfo  {journal} {Phys. Rev. B}\ }\textbf {\bibinfo {volume} {92}},\ \bibinfo {pages} {245418} (\bibinfo {year} {2015}{\natexlab{b}})}\BibitemShut {NoStop}%
\bibitem [{\citenamefont {Goswami}\ and\ \citenamefont {Harbola}(2013)}]{PhysRevA.88.013842_hpg01}%
  \BibitemOpen
  \bibfield  {author} {\bibinfo {author} {\bibfnamefont {H.~P.}\ \bibnamefont {Goswami}}\ and\ \bibinfo {author} {\bibfnamefont {U.}~\bibnamefont {Harbola}},\ }\href {\doibase 10.1103/PhysRevA.88.013842} {\bibfield  {journal} {\bibinfo  {journal} {Phys. Rev. A}\ }\textbf {\bibinfo {volume} {88}},\ \bibinfo {pages} {013842} (\bibinfo {year} {2013})}\BibitemShut {NoStop}%
\bibitem [{\citenamefont {Dorfman}\ \emph {et~al.}(2013)\citenamefont {Dorfman}, \citenamefont {Voronine}, \citenamefont {Mukamel},\ and\ \citenamefont {Scully}}]{doi:10.1073/pnas.1212666110Photosynthetic_reaction_center_as_a_qhe}%
  \BibitemOpen
  \bibfield  {author} {\bibinfo {author} {\bibfnamefont {K.~E.}\ \bibnamefont {Dorfman}}, \bibinfo {author} {\bibfnamefont {D.~V.}\ \bibnamefont {Voronine}}, \bibinfo {author} {\bibfnamefont {S.}~\bibnamefont {Mukamel}}, \ and\ \bibinfo {author} {\bibfnamefont {M.~O.}\ \bibnamefont {Scully}},\ }\href {\doibase 10.1073/pnas.1212666110} {\bibfield  {journal} {\bibinfo  {journal} {Proceedings of the National Academy of Sciences}\ }\textbf {\bibinfo {volume} {110}},\ \bibinfo {pages} {2746} (\bibinfo {year} {2013})},\ \Eprint {http://arxiv.org/abs/https://www.pnas.org/doi/pdf/10.1073/pnas.1212666110} {https://www.pnas.org/doi/pdf/10.1073/pnas.1212666110} \BibitemShut {NoStop}%
\bibitem [{\citenamefont {Rahav}\ \emph {et~al.}(2012{\natexlab{b}})\citenamefont {Rahav}, \citenamefont {Harbola},\ and\ \citenamefont {Mukamel}}]{PhysRevA.86.043843_Rahav_Reducued_DM_1}%
  \BibitemOpen
  \bibfield  {author} {\bibinfo {author} {\bibfnamefont {S.}~\bibnamefont {Rahav}}, \bibinfo {author} {\bibfnamefont {U.}~\bibnamefont {Harbola}}, \ and\ \bibinfo {author} {\bibfnamefont {S.}~\bibnamefont {Mukamel}},\ }\href {\doibase 10.1103/PhysRevA.86.043843} {\bibfield  {journal} {\bibinfo  {journal} {Phys. Rev. A}\ }\textbf {\bibinfo {volume} {86}},\ \bibinfo {pages} {043843} (\bibinfo {year} {2012}{\natexlab{b}})}\BibitemShut {NoStop}%
\bibitem [{\citenamefont {Harbola}\ \emph {et~al.}(2012)\citenamefont {Harbola}, \citenamefont {Rahav},\ and\ \citenamefont {Mukamel}}]{Harbola_2012Reducued_DM_2}%
  \BibitemOpen
  \bibfield  {author} {\bibinfo {author} {\bibfnamefont {U.}~\bibnamefont {Harbola}}, \bibinfo {author} {\bibfnamefont {S.}~\bibnamefont {Rahav}}, \ and\ \bibinfo {author} {\bibfnamefont {S.}~\bibnamefont {Mukamel}},\ }\href {\doibase 10.1209/0295-5075/99/50005} {\bibfield  {journal} {\bibinfo  {journal} {Europhysics Letters}\ }\textbf {\bibinfo {volume} {99}},\ \bibinfo {pages} {50005} (\bibinfo {year} {2012})}\BibitemShut {NoStop}%
\bibitem [{\citenamefont {Giri}\ and\ \citenamefont {Goswami}(2019{\natexlab{a}})}]{PhysRevE.99.022104_giri_ml}%
  \BibitemOpen
  \bibfield  {author} {\bibinfo {author} {\bibfnamefont {S.~K.}\ \bibnamefont {Giri}}\ and\ \bibinfo {author} {\bibfnamefont {H.~P.}\ \bibnamefont {Goswami}},\ }\href {\doibase 10.1103/PhysRevE.99.022104} {\bibfield  {journal} {\bibinfo  {journal} {Phys. Rev. E}\ }\textbf {\bibinfo {volume} {99}},\ \bibinfo {pages} {022104} (\bibinfo {year} {2019}{\natexlab{a}})}\BibitemShut {NoStop}%
\bibitem [{\citenamefont {Giri}\ and\ \citenamefont {Goswami}(2019{\natexlab{b}})}]{PhysRevE.99.022104_2019_hpg_ml}%
  \BibitemOpen
  \bibfield  {author} {\bibinfo {author} {\bibfnamefont {S.~K.}\ \bibnamefont {Giri}}\ and\ \bibinfo {author} {\bibfnamefont {H.~P.}\ \bibnamefont {Goswami}},\ }\href {\doibase 10.1103/PhysRevE.99.022104} {\bibfield  {journal} {\bibinfo  {journal} {Phys. Rev. E}\ }\textbf {\bibinfo {volume} {99}},\ \bibinfo {pages} {022104} (\bibinfo {year} {2019}{\natexlab{b}})}\BibitemShut {NoStop}%
\bibitem [{\citenamefont {Esposito}\ \emph {et~al.}(2009{\natexlab{b}})\citenamefont {Esposito}, \citenamefont {Harbola},\ and\ \citenamefont {Mukamel}}]{esposito2009nonequilibrium}%
  \BibitemOpen
  \bibfield  {author} {\bibinfo {author} {\bibfnamefont {M.}~\bibnamefont {Esposito}}, \bibinfo {author} {\bibfnamefont {U.}~\bibnamefont {Harbola}}, \ and\ \bibinfo {author} {\bibfnamefont {S.}~\bibnamefont {Mukamel}},\ }\href {\doibase 10.1103/RevModPhys.81.1665} {\bibfield  {journal} {\bibinfo  {journal} {Rev. Mod. Phys.}\ }\textbf {\bibinfo {volume} {81}},\ \bibinfo {pages} {1665} (\bibinfo {year} {2009}{\natexlab{b}})}\BibitemShut {NoStop}%
\bibitem [{\citenamefont {Ptaszy\ifmmode~\acute{n}\else \'{n}\fi{}ski}(2018)}]{PhysRevB.98.085425_krytz}%
  \BibitemOpen
  \bibfield  {author} {\bibinfo {author} {\bibfnamefont {K.}~\bibnamefont {Ptaszy\ifmmode~\acute{n}\else \'{n}\fi{}ski}},\ }\href {\doibase 10.1103/PhysRevB.98.085425} {\bibfield  {journal} {\bibinfo  {journal} {Phys. Rev. B}\ }\textbf {\bibinfo {volume} {98}},\ \bibinfo {pages} {085425} (\bibinfo {year} {2018})}\BibitemShut {NoStop}%
\bibitem [{\citenamefont {Pietzonka}\ and\ \citenamefont {Seifert}(2018)}]{PhysRevLett.120.190602_seifert}%
  \BibitemOpen
  \bibfield  {author} {\bibinfo {author} {\bibfnamefont {P.}~\bibnamefont {Pietzonka}}\ and\ \bibinfo {author} {\bibfnamefont {U.}~\bibnamefont {Seifert}},\ }\href {\doibase 10.1103/PhysRevLett.120.190602} {\bibfield  {journal} {\bibinfo  {journal} {Phys. Rev. Lett.}\ }\textbf {\bibinfo {volume} {120}},\ \bibinfo {pages} {190602} (\bibinfo {year} {2018})}\BibitemShut {NoStop}%
\bibitem [{\citenamefont {Saryal}\ and\ \citenamefont {Agarwalla}(2021)}]{PhysRevE.103.L060103_saryal}%
  \BibitemOpen
  \bibfield  {author} {\bibinfo {author} {\bibfnamefont {S.}~\bibnamefont {Saryal}}\ and\ \bibinfo {author} {\bibfnamefont {B.~K.}\ \bibnamefont {Agarwalla}},\ }\href {\doibase 10.1103/PhysRevE.103.L060103} {\bibfield  {journal} {\bibinfo  {journal} {Phys. Rev. E}\ }\textbf {\bibinfo {volume} {103}},\ \bibinfo {pages} {L060103} (\bibinfo {year} {2021})}\BibitemShut {NoStop}%
\end{thebibliography}%


%

\end{document}